\documentclass{aa}
\pdfoutput=1

\usepackage{graphicx,amsmath}
\usepackage{txfonts}
\usepackage{latexsym}
\usepackage{graphics}
\usepackage{afterpage}
\usepackage[]{hyperref}
\usepackage{dcolumn}
\usepackage{longtable}
\usepackage{booktabs}
\usepackage{subcaption}%
\usepackage[export]{adjustbox}

\usepackage{units}

\makeatletter
 \def\hlinewd#1{%
   \noalign{\ifnum0=`}\fi\hrule \@height #1 \futurelet
    \reserved@a\@xhline}
\makeatother

\begin{document} 

   \title{
     A galaxy cluster in the innermost Zone of Avoidance, \\ close to the radio phoenix VLSS\,J2217.5+5943 
     }

   \author{
     W. Kollatschny \inst{1}, 
     H. Meusinger \inst{2},
     M. Hoeft\inst{2},
     G. J. Hill \inst{3},  
     M.W. Ochmann \inst{1,4}, 
     G. Zeimann \inst{5},
     D. Froebrich \inst{6},     
     and     
     S. Bhagat  \inst{2}
     }

   \institute{
          Institut f\"ur Astrophysik, Universit\"at G\"ottingen,
          Friedrich-Hund Platz 1, 37077 G\"ottingen, Germany
          \and
          Th\"uringer Landessternwarte, Sternwarte 5, 07778 Tautenburg, Germany 
          \and
          Department of Astronomy and McDonald Observatory, University of Texas at Austin, 
          2515 Speedway, Austin, TX 78712, USA
          \and
          Astronomisches Institut, Ruhr-Universit\"at Bochum,
          Universit\"atsstrasse 150, 44801 Bochum, Germany
          \and
          Hobby Eberly Telescope, University of Texas, Austin, Austin, TX, 78712, USA
          \and
          Centre for Astrophysics and Planetary Science, School of Physical Sciences,
          University of Kent, Canterbury CT2 7NH, UK        
          }
          
  \date{Received April 12, 2021; accepted May 17, 2021}

 \abstract
   { % Context 
   Galaxy clusters grow by mergers with other clusters and groups. 
   Extended regions of diffuse radio emission with a steep radio spectral index are thought to be indicators of such merger events. 
   Extended radio sources with a significantly curved spectrum and complex morphology have been found in several galaxy clusters. 
   It has been proposed that these so-called radio phoenices are witnesses of cluster mergers and of the presence of Active Galactic Nuclei (AGN) prior to the merger. 
   Actually, shock fronts or turbulence induced by the mergers are believed to re-energise plasma emitted in the past active phase of a galaxy.
   } { % Aims 
   The steep spectrum radio source VLSS\,J2217.5+5943 shows a complex, filamentary morphology and a curved spectrum.  Therefore, the  source has previously been classified as a radio phoenix.  However, no galaxy cluster associated with this radio source has been confidently detected so far because the source is located in the direction of the innermost zone of the Galactic Plane at $b = +2.4\degr$ (innermost Zone of Avoidance, ZoA). 
   The main aim of this work is to identify galaxies which are part of a cluster at the location of VLSS\,J2217.5+5943, determine their redshifts, and analyse their connection with the radio source.  The confirmation of a cluster would corroborate the classification of the radio source as a radio phoenix and would demonstrate that extended, diffuse radio sources are useful indicators for the presence of a galaxy cluster, in particular in the innermost ZoA. 
   } { % Methods 
   We analysed archival observations in the near infrared (UKIDSS) and mid infrared (Spitzer) to select the galaxies in the immediate neighbourhood of the radio source.
   A sample of 23 galaxies was selected as candidate cluster members.
   Furthermore, we carried out deep integral field spectroscopy covering 6450 to 10500\,\AA\ with the red unit of the Hobby-Eberly Telescope second generation low resolution spectrograph (LRS2-R).
   %A sample of 23 galaxies was selected from archival observations in the near infrared (UKIDSS) and mid infrared (Spitzer). 
   We also reanalysed archival GMRT observations at 325 and 610 MHz.
   } { % Results
   We selected 23 galaxies within a radius of 2.5 arcmin, centered on  RA=$22^{\rm h} 17\fm5$, DE=$+59\degr 43\arcmin$ (J2000).
   Spectra were obtained for three of the brightest galaxies.
   For two galaxies we derived redshifts of $z = 0.165$ and $z = 0.161$, based on NaD absorption and TiO band heads. 
   Their spectra correspond to E-type galaxies. 
   Both galaxies are spatially associated with VLSS\,J2217.5+5943. 
   The spectrum of the third galaxy, which is slightly more distant from the radio source, indicates a LINER at  $z = 0.042$.
   It is apparently a foreground galaxy with respect to the cluster we identified. 
   } { % Conclusion 
   VLSS\,J2217.5+5943 is associated with a massive galaxy cluster at redshift $z = 0.163\pm .003$, supporting its classification as radio phoenix. 
   The intrinsic properties of the radio source, computed for the cluster redshift, are in good agreement with those of other known radio phoenices. 
   The identification of the galaxy cluster demonstrates, that far-red spectroscopy with LRS2-R succeeds in determining the redshift of galaxies in the innermost ZoV. 
   Moreover, it confirms that radio sources can be useful indicators for the presence of galaxy clusters in the ZoA. 
} 
% 5 {} token are mandatory


  \titlerunning{Galaxy cluster around 24P73}
  \authorrunning{W. Kollatschny et al.}

   \maketitle
%
%

%**********************************************************************************
%
\section{Introduction}\label{sect:intro}
%
%**********************************************************************************

Galaxies are arranged in groups, clusters, superclusters, and large scale structures such as filaments and walls.
One challenge for tracing extragalactic structures is the strong extinction of stellar light from the ultraviolet to the infrared caused by interstellar dust particles in our Galaxy. 
The strong concentration of the interstellar dust towards low Galactic latitudes $|b| < 10\deg$ leads to the well-known phenomenon of the so-called Zone of Avoidance (ZoA) \citep[e.g.,][]{Kraan_2005}. 
For example, it has been demonstrated that the Milky Way disk obscures the central part of the dynamically important Great Attractor region \citep{Dressler_1987} as well as the Perseus-Pisces Supercluster \citep{Giovanelli_1982}.
In the recent years, more comprehensive catalogues of galaxies in the ZoA were produced: 
the 2MASS redshift survey \citep[2MRS;][]{Macri_2019, Lambert_2020} and the Zone of Avoidance catalogue of 2MASS bright galaxies (see \cite{Schroeder_2019b} and references therein).
The 2MRS aims to map the three-dimensional distribution of an all-sky flux-limited sample of \textasciitilde 45\,000 galaxies. 
Nevertheless, redshifts are missing for \textasciitilde 18\,\% of the catalogue entries that are located within the ZoA and nothing is known about galaxies in the innermost ZoA defined as $|b| < 8\degr$ for $-30\degr < l < 30\degr$ and $|b| < 5\degr$ otherwise 
\citep{Macri_2019}.
Further catalogues are based on the 21\,cm HI emission line:
the Nancay HI Zone of Avoidance survey of 2MASS bright galaxies \citep{Kraan_2018}, and a catalogue of EBHIS HI-detected galaxies in the northern ZoA \citep{Schroeder_2019a}.
However, still very few galaxy redshifts are known for objects at lowest Galactic latitudes for $|b| < 5\degr$
\citep[see Fig. 12 in][]{Lambert_2020}.

Galaxy clusters in the ZoA have also been searched for in the X-ray band by identifying candidates in the ROSAT All-Sky Survey and related galaxies were then targeted in spectroscopic follow-up observations \citep{Kocevski_2007}. 
The authors report 57 galaxy clusters with $|b| < 20\degr$ in the second flux-limited `Clusters in the Zone of Avoidance' survey (CIZA) cluster catalogue, 13 of them in the ZoA as defined above, and four of them in the innermost ZoA.  

Even though radio waves are much less affected by interstellar dust, the investigation of extragalactic radio sources in the innermost ZoA is greatly hampered by the difficulty to identify optical counterparts (and thus obtaining redshifts).
The radio source 24P73, comprising the extended emission VLSS\,J2217.5+5943 at $l = 104\fdg6, b = 2\fdg4$, was first described in detail by \citet{Green_1994} based on follow-up observations of a few promising sources taken from the second Dominion Radio Astrophysical Observatory survey of the northern Galactic Plane at 408\,MHz and 1.42\,GHz \citep{Higgs_1989}. 24P73 was selected because of its unusually steep radio spectrum.  
\citet{Green_1994} pointed out that 24P73 is similar to another ultra steep spectrum source, 26P218, which is also at low Galactic latitude. 
These authors suggested that both 24P73 and 26P218 might be of extragalactic origin, namely radio halos or relics located in galaxy clusters, rather than being galactic pulsars as suggested before. 
On the other hand, they found it surprising to find two such rare ultra steep spectrum sources close to the Galactic Plane. 
The subsequent study by \citet{Green_2004} confirmed that 26P218 (= WKB\,0314+57.8) is most likely related to a galaxy cluster at $z \approx 0.8$, reddened by about 6 mag of visual extinction in the ZoA. 
However, the existence of a galaxy cluster related to 24P73 could not be verified so far. 

VLSS\,J2217.5+5943 was included in a study of 26 diffuse steep spectrum sources by \citet{vanWeeren_2009} that presented 610\,MHz observations with the Giant Metrewave Radio Telescope (GMRT). 
The main aim of this project was to determine the morphology of the sources and search for diffuse structures 
which could be associated with shock fronts or turbulence in clusters. 
These authors confirmed that VLSS\,J2217.5+5943 has an extremely steep spectrum at rest-frequencies higher than about 300 MHz and
a flatter spectrum at lower frequencies. 
Moreover, they suggested that the source is a so-called radio `phoenix', i.e. a diffuse radio source where the particle acceleration is thought to be due to a large shock wave adiabatically compressing fossil plasma emitted by an AGN and still comprising mildly relativistic electrons. 
However, CIZA did not report any galaxy cluster at the location of 24P73 and no optical galaxies could be identified.
In a subsequent study, \citet{vanWeeren_2011} followed up with radio continuum observations at\,325 MHz and 1.4\,GHz and optical imaging with the William Herschel Telescope which revealed several faint red galaxies. 
Using the Hubble-R relation, a redshift of  $z = 0.15 \pm 0.1$ was estimated.
The authors confirmed the classification of VLSS\,J2217.5+5943 as a radio `phoenix'. 
They also emphasised that deep near-infrared imaging will be necessary to confirm the presence of a cluster.
Recently, Chandra observations of VLSS\,J2217.5+5943 revealed extended diffuse X-ray emission (Mandal et al., submitted to A\&A), which provides further evidence for the presence of a galaxy group or cluster.

In the present study, we used archival imaging observations in the near infrared (NIR) and mid infrared (MIR) to select galaxies in the field of VLSS\,J2217.5+5943. 
Of course, spectroscopic redshifts are of crucial importance for the interpretation of that galaxy sample. 
Despite the enormous advances in exploring galaxies in the ZoA, no spectroscopic redshifts of galaxies were reported so far for this sky region.  
We present spectra of three galaxies obtained with  
%the red unit of 
the new second generation low resolution spectrograph \citep[LRS2;][]{chonis_2016} on the upgraded 10\,m Hobby-Eberly Telescope \citep[HET;][]{hill_2018}. 
The paper is structured as follows. In Sect.\,2, we describe the imaging observations and the construction of the galaxy sample. 
In Sect.\,3, we present the spectroscopic observations and data analysis. In Sect.\,4, we present the radio data analysis. Section 5 is devoted to the description of the galaxy cluster, including a rough estimate of the cluster mass.  The results are summarised in Sect.\,6.
We assume $\Lambda$CDM cosmology with $H_0$~=~73~km s$^{-1}$ Mpc$^{-1}$, $\Omega_{\Lambda}$~=~0.73 and $\Omega_{\rm M}$~=~0.27.

%**********************************************************************************
%
\section{Imaging}\label{sect:imaging}
%
%**********************************************************************************

\subsection{Selection of galaxies from UGPS}\label{sect:UGPS}

\begin{table}[htbp]
\caption{Positions, observed Petrosian K magnitudes, J-K colours, and absolute K magnitudes  for the 23 selected galaxies. 
%**Explain the source of K, J-K values**
}
\begin{tabular}{rccccc} 
\hline\hline 
\noalign{\smallskip}
  G &RA J2000.0&DE J2000.0& $K$ &$J-K$& $M_{\rm K}$   \\
\hline                    
\noalign{\smallskip}
  1 & 22 17 15.7 & 59 41 57 & 16.27 & 1.56 & -23.31  \\
  2 & 22 17 17.2 & 59 43 22 & 14.00 & 1.65 & -22.64  \\
  3 & 22 17 20.4 & 59 41 54 & 15.61 & 2.23 & -23.94  \\
  4 & 22 17 21.8 & 59 41 28 & 17.48 &    *  & -22.44  \\
  5 & 22 17 21.9 & 59 43 40 & 15.60 & 2.53 & -24.04  \\
  6 & 22 17 23.3 & 59 41 51 & 14.06 & 2.22 & -25.49  \\
  7 & 22 17 23.6 & 59 42 37 & 16.02 & 2.53 & -23.62  \\
  8 & 22 17 24.4 & 59 41 39 & 15.97 & 1.98 & -23.57  \\
  9 & 22 17 25.4 & 59 42 13 & 16.24 & 2.50 & -23.39  \\
 10 & 22 17 26.5 & 59 42 35 & 14.23 & 1.90 & -25.31  \\
 11 & 22 17 28.4 & 59 43 06 & 15.44 & 2.46 & -24.17  \\
 12 & 22 17 29.0 & 59 44 29 & 16.96 &    *  & -22.95  \\
 13 & 22 17 29.9 & 59 42 34 & 14.06 & 2.02 & -25.48  \\
 14 & 22 17 30.4 & 59 43 28 & 14.67 & 1.94 & -24.87  \\
 15 & 22 17 30.7 & 59 42 49 & 15.78 & 2.45 & -23.83  \\
 16 & 22 17 30.9 & 59 43 29 & 14.40 & 2.13 & -25.14  \\
 17 & 22 17 34.5 & 59 40 51 & 14.76 & 2.66 & -24.95  \\
 18 & 22 17 36.2 & 59 42 51 & 16.20 & 3.01 & -23.81  \\
 19 & 22 17 37.5 & 59 43 38 & 16.24 & 1.74 & -23.31  \\
 20 & 22 17 38.1 & 59 43 43 & 15.55 & 3.32 & -24.90  \\
 21 & 22 17 41.4 & 59 41 01 &     *   &    *  &      *  \\
 22 & 22 17 44.5 & 59 42 23 & 16.38 & 2.05 & -23.16  \\
 23 & 22 17 45.2 & 59 43 58 & 15.53 & 1.96 & -24.01  \\
 \hline        
\end{tabular}                             
\label{tab:galaxies}                    
\end{table}  
% /2020/photometry/comp_abs_mags_galaxies_23.prg,  Eingabedatei: 23_UGPS_galaxies_final
% h = 0.73, Omega_m = 0.27, Omega_L = 0.73

We extracted NIR images in J, H, and K of the field around 24P73 from the Galactic Plane Survey \citep[UGPS,][]{Lucas_2008} of the United Kingdom Infrared Deep Sky Survey  \citep[UKIDSS,][]{Lawrence_2007}. 
The survey utilised the Wide Field Camera \citep[WFCAM,][]{Casali_2007} on the United Kingdom Infrared Telescope (UKIRT).  
The photometric system and the  science archive are described in \citet{Hewett_2006} and  \citet{Hambly_2008}.
UGPS observations were conducted with exposure times of 80, 80 and 40\,s in J, H, and K, respectively. This leads to typical 5$\sigma$ detection limits of  $J= 19.7, H= 19.0,$ and $K= 18.05$ \citep{Warren_2007}. The 0.4 arcsec pixel size of WFCAM was combined with micro-stepping observations, leading to 0.2 arcsecond pixels in the final images, with typical seeing being better than 0.8 arcseconds. 
 This combination of depth and resolution markedly improved on 2MASS sensitivity \citep{Skrutskie_2006}, especially in the crowded fields near the Galactic Plane.   The astrometry is anchored to 2MASS and the typical precision for unblended bright stars is 0.1 arcsec r.m.s. \citep{Dye_2006}.  The images of the field of  24P73 in H and K where taken on 2011.10.24 and the J images in part on that day and on 2011.10.23.  The seeing in the images is of the order of 0.9 arcsec. The UGPS images of this field were downloaded from the WSA database\footnote{http://wsa.roe.ac.uk//index.html}.  
All data has been stacked to obtain the deep images shown in the work, using the  image mosaic engine Montage\footnote{http://montage.ipac.caltech.edu/}.

% Fig.01
\begin{figure*}[htbp]
\includegraphics[width=18.0cm,angle=0]{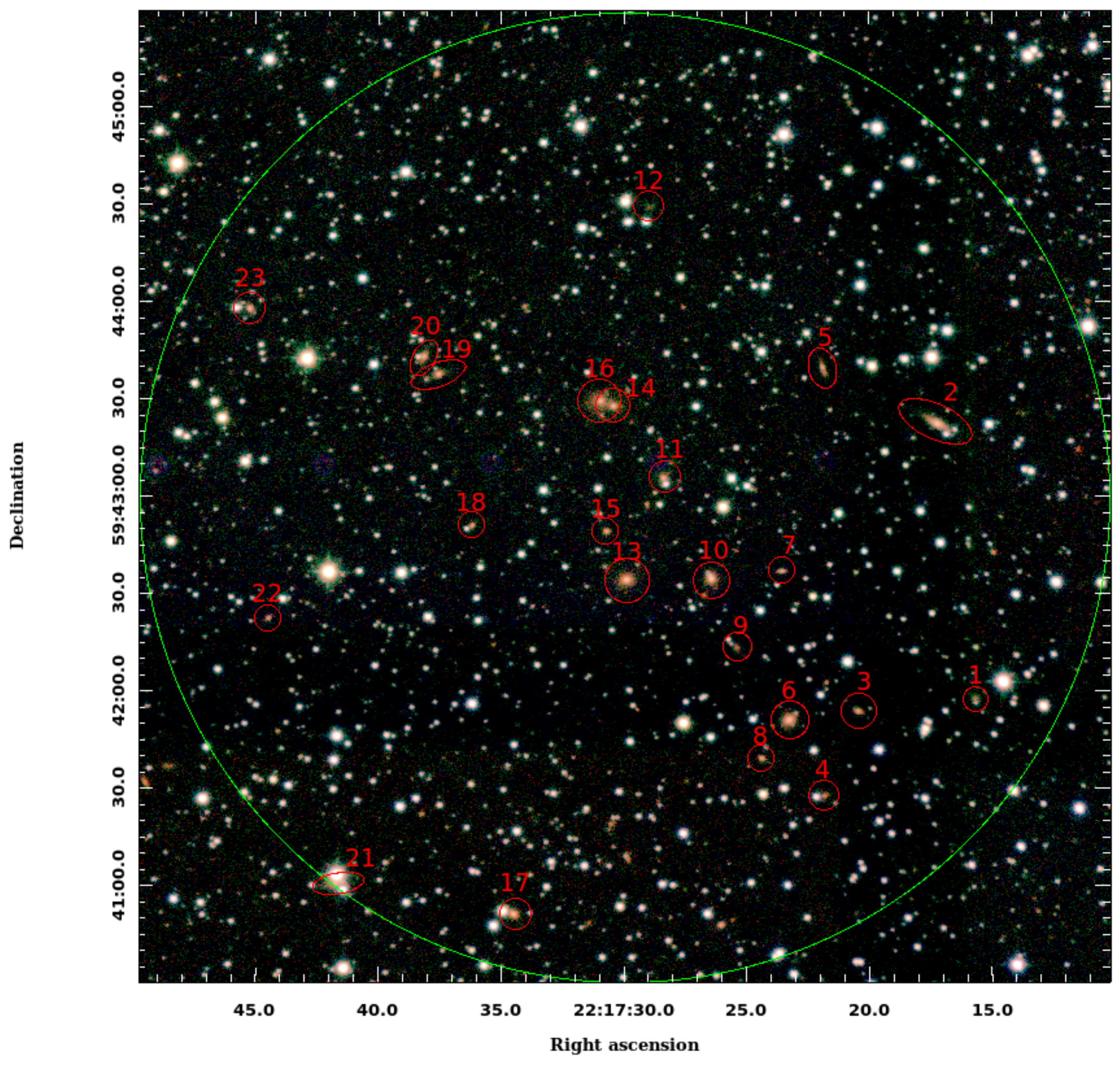}
\caption{RGB composite of the central $5\arcmin \times 5\arcmin$ field around the position of the radio source 24P73 ($22^{\rm h} 17\fm5 +59\degr 43\arcmin$, J2000) from the UGPS J, H, and K images. 
The green circle marks the selection area. Selected galaxies are indicated by their red colour and numbered in order of increasing RA.}
\label{fig:UGPS_RGB}
\end{figure*}
% ~/Dokumente/GCs/VLSS2217/2016/images/crea_RGB_with_contours_and_grid.txt

\begin{figure}[htbp]
\includegraphics[viewport= 0 35 570 800,width=6.8cm,angle=270]{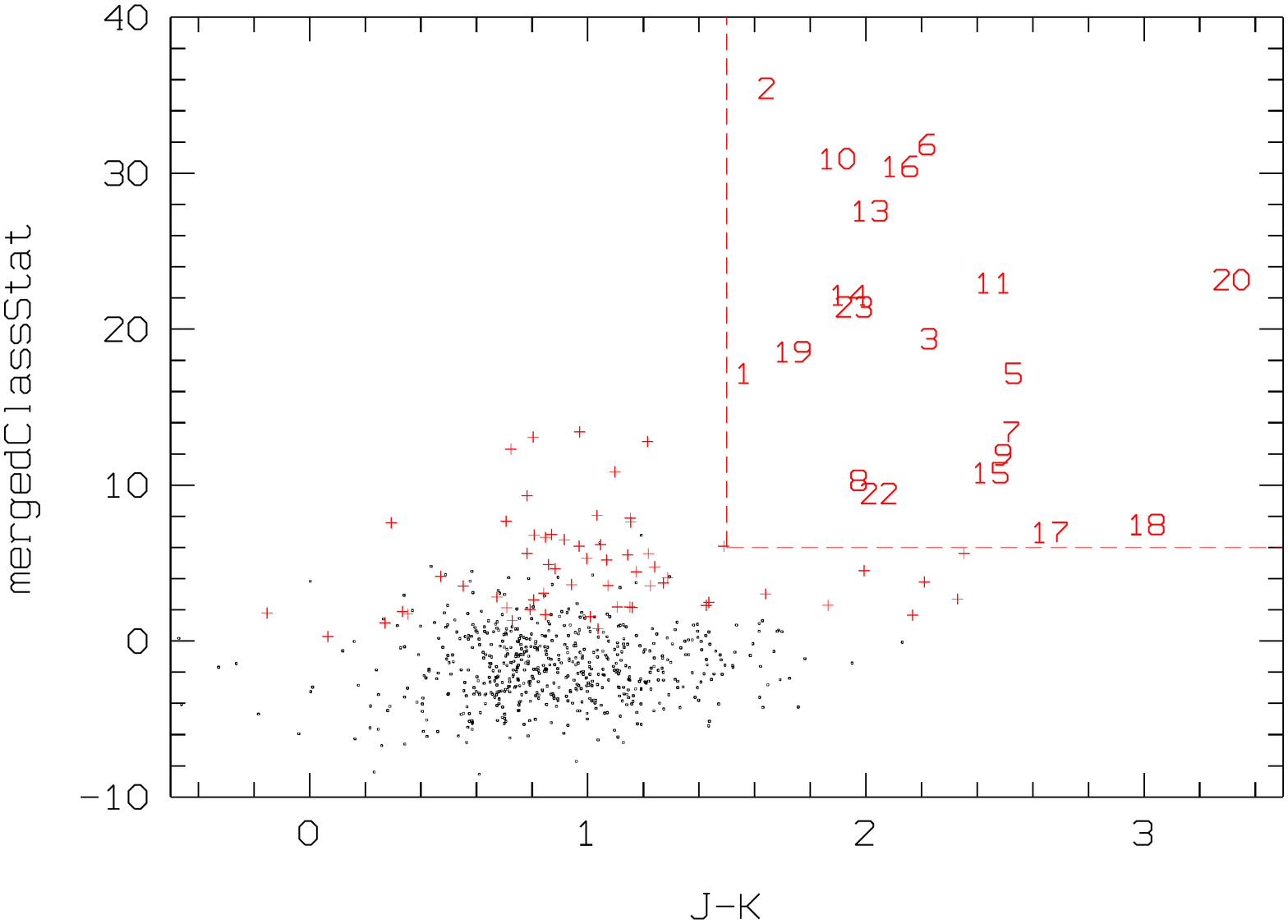}
\caption{Morphology-colour diagram for 658 UKIDSSDR11\-PLUS sources with $J<19.5$ and $K<19.5$ in the 24 square arcmin field around VLSS\,J2217.5+5943. 
The colour index $J-K$ is computed from the Petrosian magnitudes.
Black and red symbols indicate sources classified as star-like or as galaxies, respectively. The dashed lines mark the selection area for  the galaxies in the present study.}
\label{fig:selected_galaxies}
\end{figure}

Figure\,\ref{fig:UGPS_RGB} shows the $5\arcmin \times 5\arcmin$ RGB colour image (red = K, green = H, blue = J) centred on 
the coordinates of the radio source 24P73
(RA $ = 22^{\rm h} 17\fm 5$,  DE $= +59\degr 43\arcmin$, J2000). The diameter of the radio source is about 2 arcmin.
The image clearly reveals a concentration of extended, 
non-stellar objects towards the centre. Furthermore, extragalactic sources beyond the screen of interstellar dust in our Galaxy are 
conspicuously indicated by their red colour caused by the strong interstellar reddening in the inner ZoA. 
In this way, combining morphological and colour information, we initially identified 16 galaxies simply by the visual inspection of Fig.\,\ref{fig:UGPS_RGB}.  
In the next step, we used a morphology-colour diagram (Fig.\,\ref{fig:selected_galaxies}) with the  UKIDSS parameter {\tt mergedClassStat} 
as an indicator of the extension of the sources. Objects classified as star-like (${\tt pStar} > 0.9$, black) in UKIDSSDR11PLUS cluster around 
{\tt mergedClassStat} $=-2$ and $J-K = 0.8$, whereas objects classified as galaxies (${\tt pGalaxy} > 0.9$, red)  have {\tt mergedClassStat} $>0$.  The visual inspection indicates that sources with {\tt mergedClassStat} $>0$ and $J-K \la 1.5$ are most likely blended stars. 
On the other hand, all galaxies from our eyeball selection were found to be 
%concentrated -- Gary: not really "concentrated", since they are spread out a lot, but rather "located"
located
in the sparsely populated top right corner of the diagram. 
Hence, we used their positions to define the locus of distant galaxies in that diagram (red dashed lines). 
Some faint objects close to the demarcation lines but outside the selection area may be galaxies of small angular size. In particular, the two objects marked as red plus signs very close to the dashed line are most likely such faint galaxies. We do not wish to extend the selection towards very small galaxies, mainly because of the increasing contamination by background galaxies that are not spatially connected to the brighter galaxies in our sample.

% Fig.02
% /2020/photometry/plot_petrocolours_all.prg

Altogether, we selected 23 extragalactic systems which constitute the galaxy sample for the present study. 
The galaxies are listed in Table\,\ref{tab:galaxies} with the data from the UKIDSSDR11PLUS database\footnote{http://wsa.roe.ac.uk:8080/wsa/menu.jsp?}.
Unfortunately, model magnitudes, as defined by the Sloan Digital Sky Survey (SDSS), are not computed by UKIDSS. 
Here, we use Petrosian magnitudes, which are known to not significantly depend on galaxy light profiles. 
The absolute magnitudes listed in the last column were computed for the assumption of $z = 0.042$ for G2 and $z = 0.163$ (Sect.\,\ref{sect:spec}) 
for all other galaxies using foreground reddening corrections with $E(B-V) = 1.8$ and  k corrections from \citet{Chilingarian_2012}.
We assume that the majority of these galaxies are  representing a coherent structure.  
The incompleteness of the galaxy sample and its `contamination'  by foreground and background galaxies are discussed in Sect.\,\ref{sect:total_mass}. 
We note that Fig.\,\ref{fig:selected_galaxies} shows only 20 galaxies in the selection area because the heavily blended galaxy G21 is missing and 
J magnitudes are unavailable for G4 and G12  in the  UKIDSSDR11PLUS database. 
It must further be noted that the catalogued magnitudes should be interpreted with caution because most galaxy images are blended by foreground stars.

\subsection{Spitzer Mapping of the Outer Galaxy}\label{sect:SMOG}

The VLSS\,J2217.5+5943 field was covered by the Spitzer Mapping of the Outer Galaxy \citep[SMOG;][]{Carey_2008} project as part of the 
Galactic Legacy Infrared Midplane Survey Extraordinaire (GLIMPSE). SMOG used the Spitzer Space Telescope \citep[SST;] []{Werner_2004} 
Infrared Array Camera \citep[IRAC;][]{Fazio_2004} in the 'Cryogenic(Cold) Mission' program phase.  
IRAC mosaic images in the four bands I1 to I4 at 3.6, 4.5, 5.8, and 8.0 $\mu$m
with a pixel size of $0\farcs6$ are available from the GLIMPSE pages\footnote{https://irsa.ipac.caltech.edu/data/SPITZER/GLIMPSE/index.html} 
at the NASA/IPAC Infrared Science Archive (IRSA). We could easily identify all galaxies from Table\,\ref{tab:galaxies} on the images in the I1 and I2 bands, whereas only the brightest galaxies are (barely) seen in the I3 and I4 bands. Figure\,\ref{fig:IRAC_Gray} shows the stack of the images from I1 and I2 for the field of the extended radio source 24P73. Compared to the UGPS image, the galaxies appear slightly larger in the IRAC images,  i.e. their extensions can be traced down to fainter surface brightness structures (see Sect.\,\ref{sect:cluster}).

\subsection{Optical surveys}\label{sect:optical}

None of the selected galaxies is clearly visible on the images from the ESO Digitised Sky Surveys (DSS) indicating that they are fainter than about 20th magnitude in the optical.  
Deeper images in optical bands are available from the Panoramic Survey Telescope and Rapid Response System \citep[Pan-STARRS or PS1;][]{Chambers_2019, Flewelling_2020} located at Haleakala Observatory, Hawaii, and from the INT/WFC Photometric H-Alpha Survey of the Northern Galactic Plane \citep[IPHAS;][]{Barentsen_2014}.

The PS1 catalogue includes measurements in five filters (grizy)$_{\rm PS1}$ covering the entire sky north of  $\delta = -30\degr$ ($3\pi$ survey) down to  $r_{\rm P1} \approx 23.3$. The data are available through the  MAST system at STScI\footnote{https://catalogs.mast.stsci.edu/panstarrs/}. With the exception of the faintest galaxies (G1, 4, 12) and the strongly blended G21, all galaxies from Table\,\ref{tab:galaxies} were identified in PS1 but relatively close to the survey limit. The bright galaxy G13, for example, is listed with an aperture magnitude $r_{\rm PS1} = 22.7 \pm 0.1$, but the Kron r magnitude was not measured.

For completeness, we also mention the 1860 square degrees imaging survey IPHAS of the Northern Milky Way  in the broad-band r, i, and 
narrow-band H$\alpha$ filters using the Wide Field Camera (WFC) on the 2.5-m Isaac Newton Telescope in La Palma. 
The observations were carried out  at a median seeing of 1.1 arcsec and to a mean 5-sigma depth of 21.2 (r), 20.0 (i) and 20.3 (H$\alpha$).
The search for the galaxies from Table\,\ref{tab:galaxies} in the IPHAS DR2 Source Catalogue \citep{Barentsen_2014} yielded a detection only for the brightest galaxy, G2. All other galaxies from Table\,\ref{tab:galaxies} have no IPHAS counterparts, neither in the catalogue nor on the images.

\subsection{Photometric redshift}\label{sect:photo_redshift}

\begin{figure}[htbp]
\includegraphics[viewport= 0 20 570 800,width=6.6cm,angle=270]{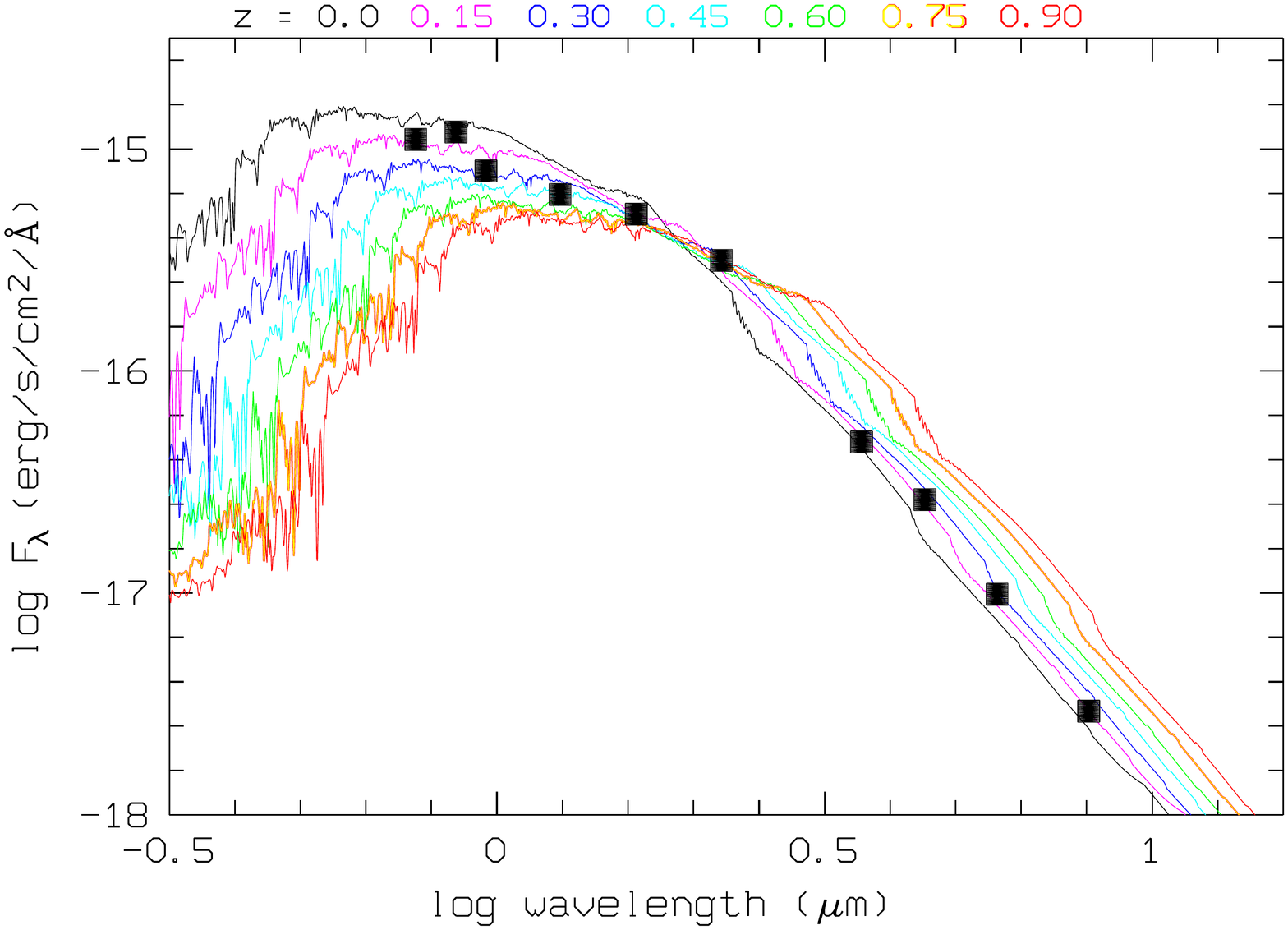}
\caption{Observed SED of G13  (black squares) based on corrected aperture magnitudes from Pan-STARRS rzy, UKIDSS JHK, and the four Spitzer  IRAC bands corrected for Galactic interstellar extinction with $E(B-V) = 2$. The formal error bars for the photometric uncertainties are smaller than the symbol size.
For comparison, the redshifted template spectrum of a 13 Gyr old E galaxy from \citet{Polletta_2007} 
is overplotted for $z = 0.0 - 0.90$ in steps of 0.15 (colours).}
\label{fig:SED_G13}
\end{figure}

The galaxy G13 is, along with G6, the second brightest in the K band. Unlike most of the other galaxies, G13 is not much blended by foreground stars  and is thus suitable for a study of its spectral energy distribution (SED). 
Figure\,\ref{fig:SED_G13} shows the SED based on broad-band magnitudes from Pan-STARRS and UGPS, 
and fluxes from Spitzer IRAC\footnote{data provided by the NASA/IPAC Extragalactic Database (NED)}. 
The difficulty of such an approach is that the different surveys provide different types of measurements for extended sources: Kron magnitudes, 
Petrosion magnitudes, and aperture magnitudes for different aperture sizes, which are not directly comparable with each other. 
To tackle this problem, we measured the radial surface brightness profile on the IRAC 1+2 image (Fig.\,\ref{fig:IRAC_Gray}) and
assumed that the profile is similar in the other bands. 
Then, this profile was used to extrapolate the measured flux in the largest aperture to the maximum size of the measured profile.

% Fig.03
% /2020/photometry/BCG/Spitzer_imaging/new_photometry/plot_WBSED_2.prg

The observed SED has to be corrected for reddening by Galactic dust, which is crucial for fields deep in the ZoA.  
The IRSA Galactic Dust Reddening and Extinction Service\footnote{https://irsa.ipac.caltech.edu/applications/DUST/} 
gives $E(B-V) = 1.30^{+0.12}_{-0.05}$ \citep{Schlafly_2011}.  However, there is a large degree of uncertainty in these values as there is
-- looking through multiple spiral arms -- the possibility of multiple dust 
temperature distributions or variable grain sizes close to the mid-plane of the Milky Way.
As noted by IRSA,\footnote{https://irsa.ipac.caltech.edu/applications/DUST/docs/background.html}
`the user of this service is advised to use caution when working deep in the Galactic Plane'.
At this point, we use  $E(B-V)$ just as a free parameter to check whether the SED is consistent with $z \approx 0.15$ from \citet{vanWeeren_2011}. 
Figure\,\ref{fig:SED_G13} shows the de-reddened SED for $E(B-V) = 2$ compared with the template SED of an 13 Gyr old E galaxy from the 
SWIRE library\footnote{http://www.sedfitting.org/Data.html} \citep{Polletta_2007} for different redshifts $z = 0 \ldots 0.9$. 
At each $z$, the template was fitted to the observed SED using the least-square method. Best agreement is found for $z \approx 0.15$ (magenta spectrum).  Of course, the resulting $z$ depends strongly on the adopted value for $E(B-V)$ with higher $z$ for smaller $E(B-V)$.

%**********************************************************************************
%
\section{Spectroscopy}\label{sect:spec}
%
%**********************************************************************************

As mentioned in Sect.\,\ref{sect:intro},
spectra and redshifts of the galaxies in the field of  24P73 are not contained in any 
%recent -Gary: not just recent, but any...
2MASS redshift surveys (2MRS)
\citep[e.g.][]{Lambert_2020}. Furthermore, they are not listed in HI-surveys of objects in the ZoA:
neither in the Nancay HI Zone of Avoidance survey of 2MASS bright galaxies \citep{Kraan_2018}, nor in the 
catalogue of EBHIS HI-detected galaxies in the northern ZoA \citep{Schroeder_2019a}.
We selected the three galaxies G2, G6, and G16, which are among to the brightest in the K band in our sample, with the aim
of obtaining spectroscopic redshifts.

% Fig.04
\begin{figure*}[htbp]
\begin{center}
\includegraphics[viewport= 0 0 730 285,width=18cm,angle=0]{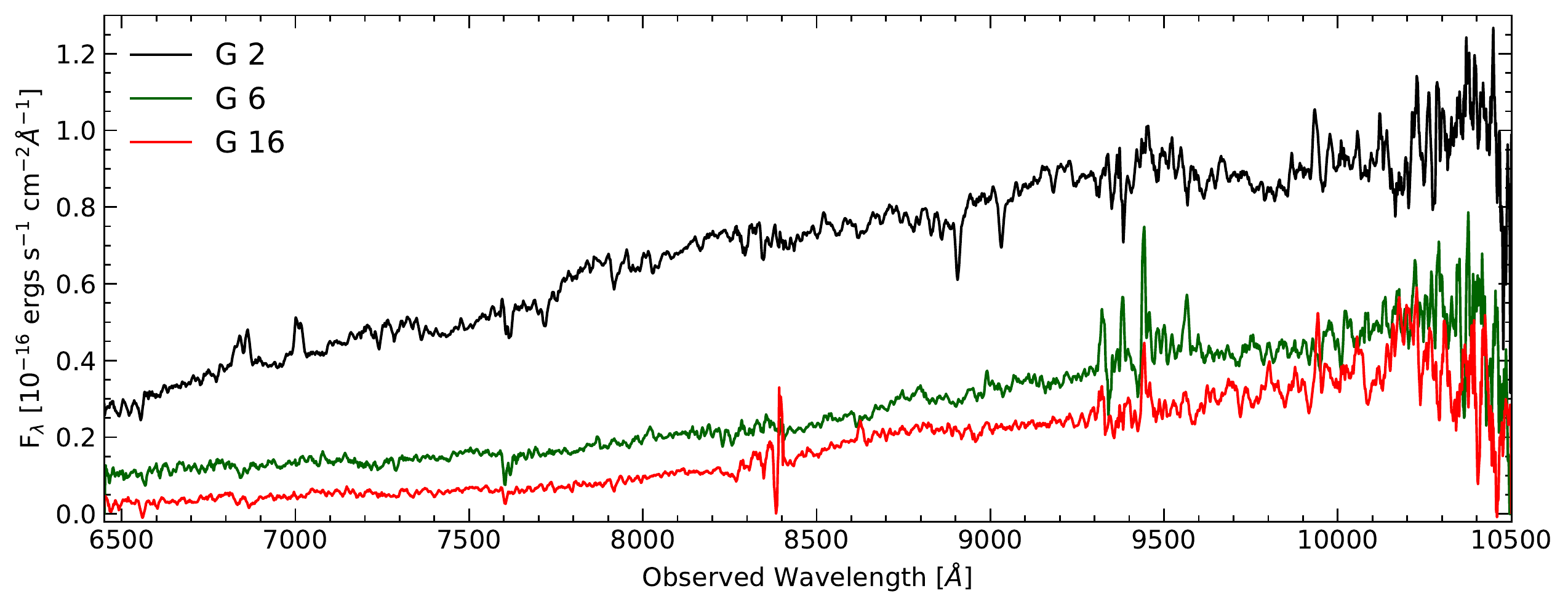}
\end{center}
\caption{Smoothed LRS2 spectra of the galaxies G2 (black), G6 (green), and G16 (red). }
\label{fig:spectra}
\end{figure*}

\subsection{Observations and data reduction}\label{sect:spec_obs}

We used the red unit of the second generation Low Resolution Spectrograph \citep[LRS2-R;][]{chonis_2016} on the upgraded HET.
%We used the Low Resolution Spectrograph LRS2-R attached at the prime focus of HET.
The HET is one of the world's largest optical telescopes,  with an aperture of 10 meters,  located at the McDonald Observatory in the Davis Mountains, west Texas, USA.  
LRS2 is an integral field spectrograph with four spectral channels covering 3700\,\AA\ to 10500\,\AA\ in separate blue and red units (LRS2-B and LRS2-R, respectively), each fed by lenslet-coupled fibre interal field units (IFUs). Each IFU covers $12\arcsec \times 6\arcsec$ with 0.6\arcsec spatial elements and full fill-factor, and couples to 280 fibres. The IFUs are separated by 100\arcsec\ on sky, but for this study we utilized only the LRS2-R unit, which simultaneously covers the wavelength range from 6450 to 8420\,\AA\ and 8180 to 10500\,\AA\ (refered to as the red and far-red channels, respectively) at a
resolution of $R \sim 2500$. This wavelength range is split into the two spectral channels with a transition at 8350\,\AA\ by a dichroic beamsplitter in the IFU.
%LRS2 is an integral field spectrograph with a $12\arcsec \times 6\arcsec$ fibre integral field unit (IFU) with lenslet
%coupling. LRS2-R simultaneously covers the wavelength range from 6450 to 8420\,\AA\ and 8180 to 10500\,\AA\ at a
%resolution of $R \sim 1800$. We simultaneously took red and far-red spectra with 
We obtained LRS2-R spectra of the three galaxies at several
epochs between June 2017 and August 2020. The spectra have been taken in blind-offset mode by means of optical reference
stars observed in the acquisition camera. Tab.\,\ref{tab:log_of_HET_obs} gives the log of our spectroscopic observations. All spectra  were
taken with identical instrumental setups and  at the same airmass owing to the particular design of the HET with fixed altitude.

The 280 fibre spectra in each of the red and far-red channels of LRS2-R were reduced with the automatic HET pipeline, Panacea\footnote{https://github.com/grzeimann/Panacea}. 
This pipeline performs basic CCD reduction tasks, wavelength calibration, fibre extraction, sky subtraction and flux
calibration. The initial sky subtraction from the automatic pipeline removes a single sky model from each fibre and although
this is adequate for many science objectives, the strong OH sky lines, spatially-extended targets, and the wavelength-dependent spectrograph resolution require a sky
residual model to identify the redshifts of our candidate galaxies. We used arc lamp exposures to build a basis of
eigenvectors for a principal component analysis of the sky residuals similar to that of \citet{Soto_2016}. We identified our
galaxies in each observation, masked a $2\farcs5$ region around the galaxies, and fit the sky residuals over the wavelengths
of strong OH emission for the remaining fibres (typically around 200 fibres). 
The best fit components were then used for the masked fibre sky residuals as well. This greatly improved the sky subtraction.
We extracted each spectrum using a 2D Gaussian model. We normalised each spectrum for a given galaxy to the mean of
the individual spectra and co-added each normalised individual spectrum.
We modelled telluric absorption using a single telluric standard star. 
The HET always observes at nearly the same elevation, allowing for a single telluric model to have greater applicability than other
telescopes. Although the telluric H$_2$O features vary from observation to observation, the O$_2$ features are more stable.

\begin{table}[htbp]
\tabcolsep+6mm
\caption{Log of spectroscopic observations.}
\centering
%\vspace{3mm}
\begin{tabular}{ccc}
\hline\hline
\noalign{\smallskip}
G     & UT Date & Exp. time (sec)\\
\hline             
\noalign{\smallskip}
  2   &       2017-08-28      &      2415   \\
       &       2017-09-22      &      2415   \\
\hline             
\noalign{\smallskip}
  6  &       2017-10-28      &      1207   \\
       &       2017-11-26      &      1207   \\
       &       2020-07-05      &      1408   \\
       &       2020-07-07      &      1409   \\
       &       2020-08-04      &      1208   \\
       &       2020-08-05      &      1208   \\   
\hline 
\noalign{\smallskip}
16   &       2017-06-16      &      1208   \\
       &       2017-10-15      &      1208   \\
\hline        
\vspace{-.7cm}
\end{tabular}
\label{tab:log_of_HET_obs}
\end{table}

\vspace{0.5cm}

\subsection{Results}\label{sect:spec_results}

Fig.\,\ref{fig:spectra} shows the smoothed observed spectra of G2 (black), G6 (green), and G16 (red). 
The smoothing was performed using a simple moving average with a box size of 15\,\AA. For further analysis, we corrected our spectra for interstellar
absorption by the Milky Way. Based on a comparison of the slope of our smoothed and de-reddened spectra of Galaxies 6 and 16 to
an elliptical template spectrum, we adopted $E(B-V) = 1.8$.

We present spectral sections around NaD and the TiO bands for G6 and G16 in Figs.\,\ref{fig:spectra_G6_G16}
and the sections around H$\alpha$ and the \ion{Ca}{ii} IR triplet for G2 in Fig.\,\ref{fig:spectra_G2}, respectively. 
As described above, the spectra have been smoothed and de-reddened. In addition, each spectral section was scaled in order to
match the spectral features (see below for more details).

\subsection{Redshift determination}\label{sect:spec_redshift}

As a first step, we tried to derive the galaxy redshifts by comparing the observed overall optical/far-red spectrum with a
redshifted template.  We used the template of an elliptical galaxy from the SDSS \citep[DR5, spectral template 24;][]{Strauss_2002}
because both the images (Figs.\,\ref{fig:UGPS_Gray}, \ref{fig:IRAC_Gray}) and the spectral characteristics indicate that G6 and G16 appear to be of type E. The resulting difference spectrum between the observed galaxy spectra and the redshifted template spectrum had to yield a smooth curve
for a the correct redshift value. Moreover, the general intensity distribution of the difference spectrum should only reproduce the
reddening by the Milky Way. In this way, we derived a first estimate of $z = 0.16 \pm 0.04$.

% Fig.05
\begin{figure}[htbp]
\includegraphics[width=1.0\linewidth,angle=0]{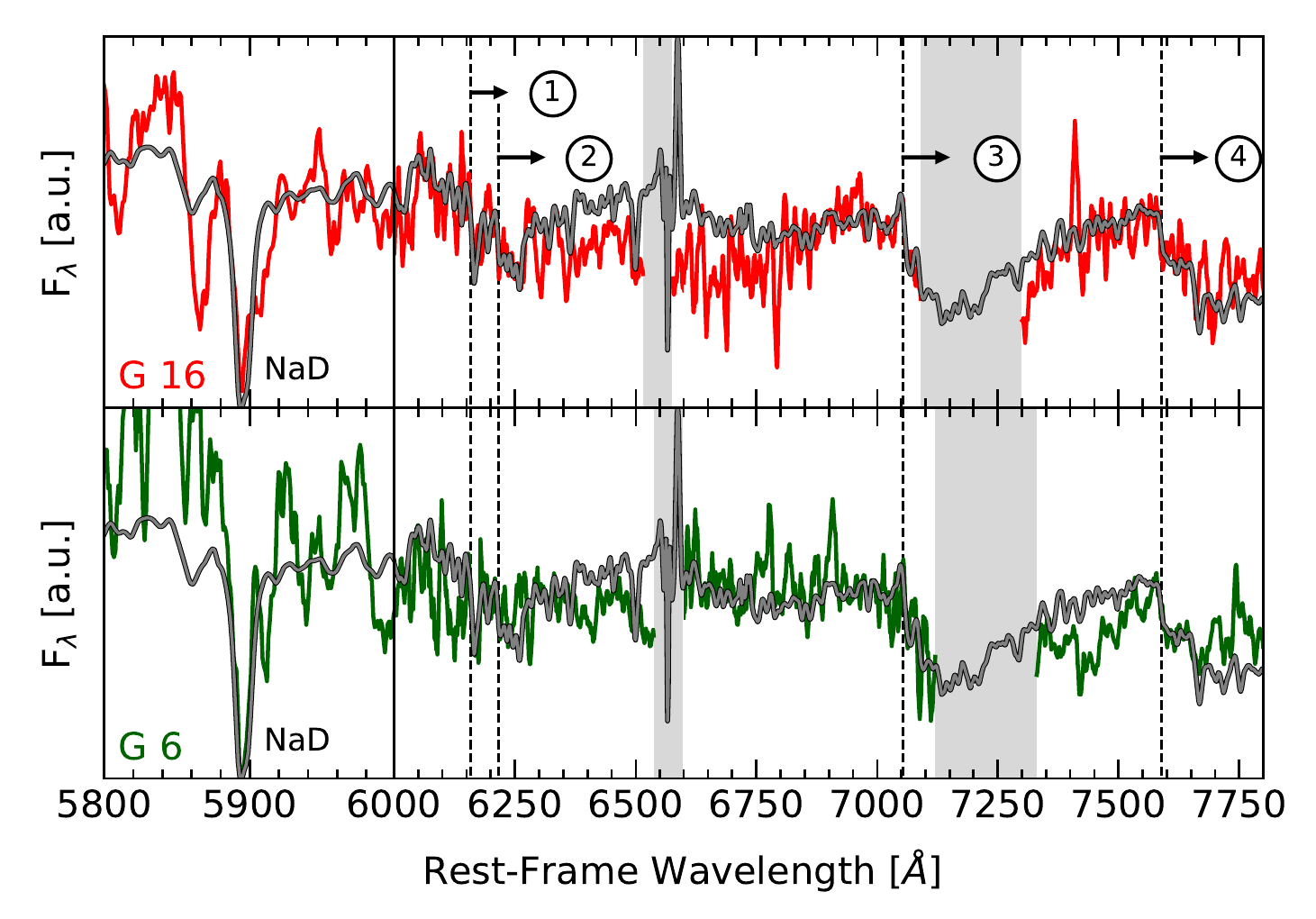}
\caption{Red/far-red spectra of galaxies G16 (red) and G6 (green) and of an elliptical template spectrum (grey) from 5800
to 7800\,\AA\ (rest-frame). The NaD absorption as well as the band heads of TiO at $\lambda$6158 (1), $\lambda$6216 (2),
$\lambda$7053 (3), and $\lambda$7589 (4) are marked. The region around the [\ion{N}{ii}] and H$\alpha$ emission complex of
the template galaxy is shaded in grey. That is because it coincides with the strong telluric A-band (7590-7660\,\AA) in the
observed frame. A region around 7200\,\AA\, corresponding to roughly 8370\,\AA\ in the observed frame, that is close to the wavelength of the dichroic filter transition between the two channels and is also contaminated
by night-sky emission lines, is also shaded in grey.}
\label{fig:spectra_G6_G16}
\end{figure}

In a second step, we derived the redshifts based on spectral features to confirm our first rough estimation. In addition
to the NaD$\lambda$5892 lines \citep[e.g.][]{Kinney_1996}, the far-red spectra of elliptical galaxies are dominated by
titanium oxide (TiO) bands from 6000 to 9000\,\AA.  These TiO bands are caused by the superposition of K5 to M9 stellar spectra.
Spectra of those types are shown e.g. in \citep{Kirkpatrick_1991}. Some of the strongest bands in the wavelength range
from 6000 to 8000\,\AA\ are the TiO$\lambda$6158,$\lambda$6216,$\lambda$7053, and $\lambda$7589 $\gamma$-bands 
\citep{Kirkpatrick_1991, Valenti_1998}. 
Their relative strength and depth are different for the individual K and M type spectra. 
We therefore compared the position of the band heads of these TiO bands -- in addition to the NaD lines -- 
with the spectral features of the template spectrum of an elliptical galaxy in order to derive the redshifts of galaxies G6 and
G16. 

To this end, we scaled the TiO band heads as well as the NaD line in the galaxy spectra such that their amplitude matches the
corresponding band head or line, respectively, in the template spectrum. This procedure resulted in four spectral sections,
namely around NaD$\lambda$5892, TiO$\lambda$6158 and $\lambda$6216, $\lambda$7053, and TiO$\lambda$7589. The boundaries
between the sections are given by the telluric O$_2$ absorption band at 7590-7660\,\AA\ and a region at around 8370\,\AA\ (both in
the observed frame) heavily contaminated by night-sky emission. In addition, we chose the boundary between NaD$\lambda$5892
and TiO$\lambda$6158 as well as $\lambda$6216 to be at 6000\,\AA. 
In contrast to G16, the amplitude scaling factors of the $\lambda$6158 and $\lambda$6216 band heads in G6 do not match. 
We therefore determined their scaling factor individually and combined both resulting spectra at 6178\,\AA. 
Figure\,\ref{fig:spectra_G6_G16} shows the composite spectra of G6 
(green line) and G16 (red line) superimposed on the SDSS template spectrum (grey lines) of an elliptical galaxy. We derived a
redshift of $z = 0.165\pm .001$ for G16 and a redshift of $z = 0.161 \pm .001$ for G6 (Table\,\ref{tab:spec_results}).

% Fig.06
\begin{figure}[htbp]
\includegraphics[width=1.0\linewidth,angle=0]{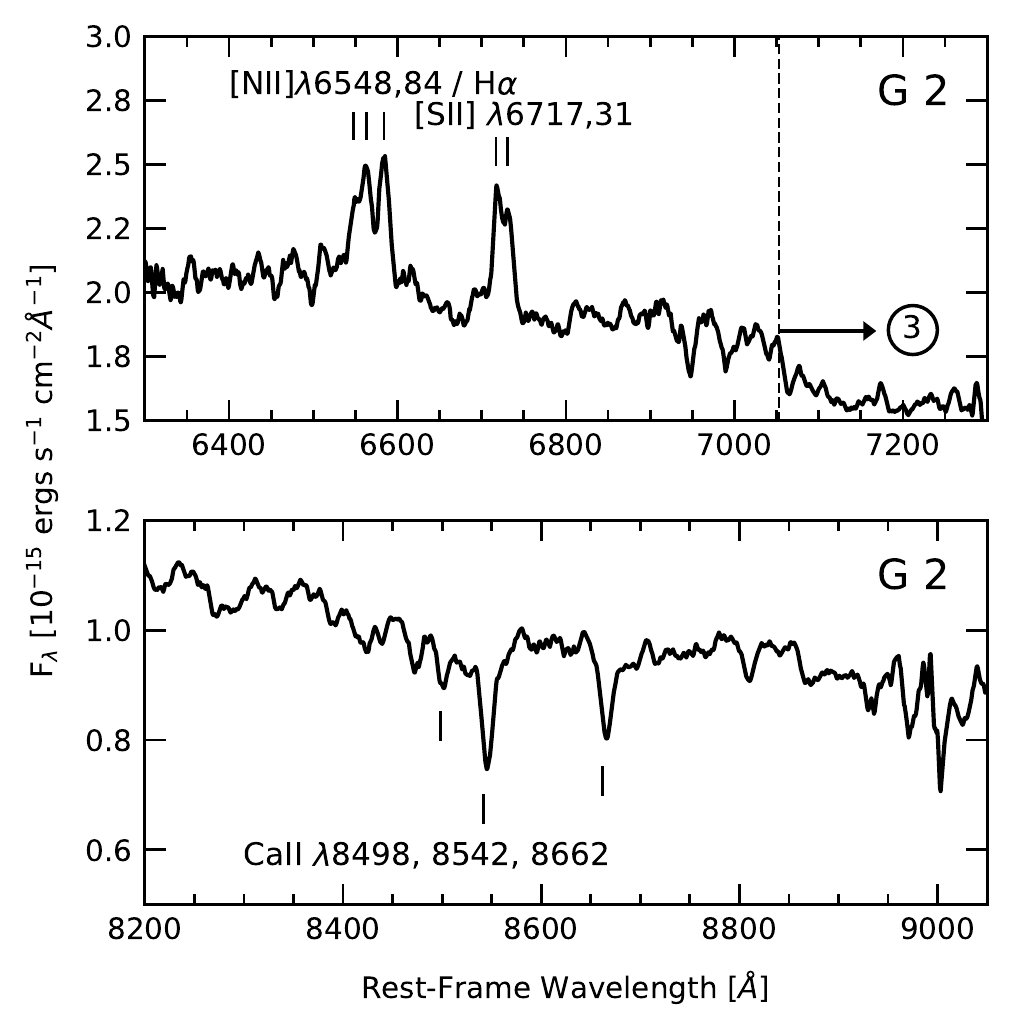}
\caption{Sections around the H$\alpha$ region and the TiO$\lambda$7053 band head
(3), as well as around the \ion{Ca}{ii} IR triplet of G2.}
\label{fig:spectra_G2}
\end{figure}

Galaxy G2 shows strong H$\alpha$, [\ion{N}{ii}]$\lambda6548,\lambda6584$, and [\ion{S}{ii}]$\lambda$6717,$\lambda$6731 emission lines 
-- and in addition the \ion{Ca}{ii} IR triplet (8498, 8542, 8662\,\AA) in absorption -- at a redshift of $z = 0.0422 \pm .0001$ (see Fig.~\ref{fig:spectra_G2}).
It does not show as strong TiO bands as in the other two cases, the strongest band head is that of the TiO$\lambda$7053 band.
This galaxy is discussed in more detail in Sect.\,\ref{sect:G2}.

\begin{table}[htbp]
%\tabcolsep+6mm
\caption{Results from the spectroscopic observations. The R magnitudes are given in  three different systems: 
AB \citep{1983ApJ...266..713O},  B98 = Bessell98 \citep{1998A&A...333..231B}, and  S96 = \citet{Stone_1996}.}
\centering
%\vspace{3mm}
\begin{tabular}{cccccc}
\hline\hline
\noalign{\smallskip}
G    &  $z$                        &    type    & \multicolumn{3}{c}{$R$}\\
\cmidrule{4-6}
       &                                &               &  AB                    & B98           &  S96 \\
\hline             
\noalign{\smallskip}
 2   & $0.0422\pm.0001$ & LINER  & 19.93                 & 19.65         & 20.20       \\
 6   & $0.161\pm.001$     &  E          & 20.97                  & 20.69         & 21.25       \\
16   & $0.165\pm.001$    &  E          & 22.33                  & 22.05         & 22.61       \\
\hline        
\vspace{-.7cm}
\end{tabular}
\label{tab:spec_results}
\end{table}

\vspace{0.5cm}

The results from  the spectroscopic observations are summarised in Table\,\ref{tab:spec_results}. 
The  given  R magnitudes were  derived  from  the  observed mean fluxes in the spectra at 6450 - 6550\,\AA. 
They have  been  calculated  also in  the STMAG system  that is  used  by  the  Hubble  Space  Telescope 
photometry packages \citep{Stone_1996}.

%**********************************************************************************
%
\section{Radio data analysis}
\label{sect:radio}
%
%**********************************************************************************

The source 24P73 has been observed with GMRT at 610\,MHz and 325\,MHz in 2008 and 2009. Total on-source time for the 610\,MHz and 325\,MHz observations was 7 hours 42 minutes and 9 hours 40 minutes, respectively. We calibrated and imaged the archival data using the Source Peeling and Atmospheric Modelling \citep[SPAM,][]{2009A&A...501.1185I}. The pipeline executes a semi-automated series of iterative flagging, calibration and imaging, most importantly, involving a procedure allowing a direction-dependent calibration of the data. The data sets for different days were combined into a single one for imaging. The initial flux and bandpass calibrations for 325 MHz and 610 MHz data sets were carried out using the standard flux calibrators 3C286 and 3C48, respectively. The resulting surface brightness distribution is shown in Fig.\ref{fig:UGPS_Gray} (green contours).  

\begin{figure}[htpb]
\begin{center}
\includegraphics[viewport= 20 20 590 820,width=6.6cm,angle=270]{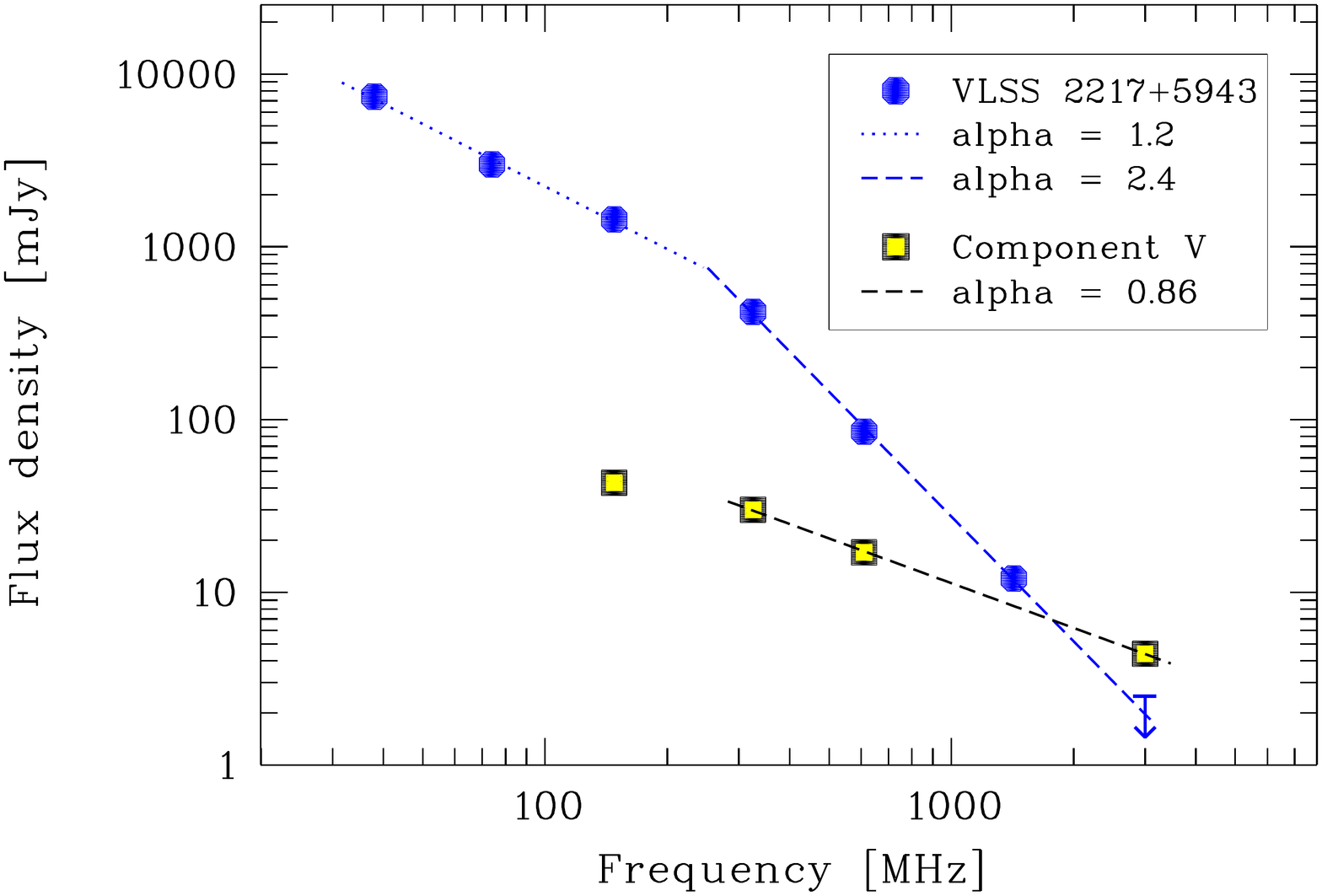}
\caption{Spectrum of VLSS\,J2217.5+5943, i.e., components II-IV (blue), and component V (dark yellow). Data are from 8C catalogue (38\,MHz), VLSS (74\,MHz), TGSS-ADR (150\,HMz), GMRT 325\,MHz and 610\,MHz, WSRT (1.4\,GHz) and VLASS (3\,GHz). The upper limit from VLASS corresponds to the 68\,\% confidence limit.  }
\label{fig:radio_spectrum}
\end{center}
\end{figure}

Additionally, we measure the flux density of the source VLSS J2217.5+5943 in the snapshot image of the Very Large Array Low-Frequency Sky Survey Redux \citep[VLSSr,][]{2014MNRAS.440..327L} and the cutout image of the TIFR GMRT Sky Survey Alternative Data Release \citep[TGSS ADR,][]{2017A&A...598A..78I}. Moreover, we include the flux values reported in 8C Revised Rees Survey 38-MHz Source catalogue \citep[8C,][]{1990MNRAS.244..233R,1995MNRAS.274..447H} and the value obtained by Westerbork Synthesis Radio Telescope (WSRT) observations as reported by \citet{vanWeeren_2009}.  Finally, we analysed Quick Look continuum images of the Very Large Array Sky Survey \citep[VLASS,][]{2020PASP..132c5001L}. 
We also determined the flux density for the compact  source J221736.2+594407 (component V, see Sect.\,\ref{sect:cluster}), where possible.  The resulting radio spectra and spectral indices $\alpha$  (for $S_\nu \propto \nu^{-\alpha}$) are shown in Fig.\,\ref{fig:radio_spectrum}.

% Fig.07
% /VLSS2217/2020/radio/plot_spectrum.prg

%**********************************************************************************
%
\section{The galaxy cluster in the field of VLSS\,J2217.5+5943}\label{sect:cluster}
%
%**********************************************************************************

\begin{figure*}[htbp]
\includegraphics[width=16.0cm,angle=0]{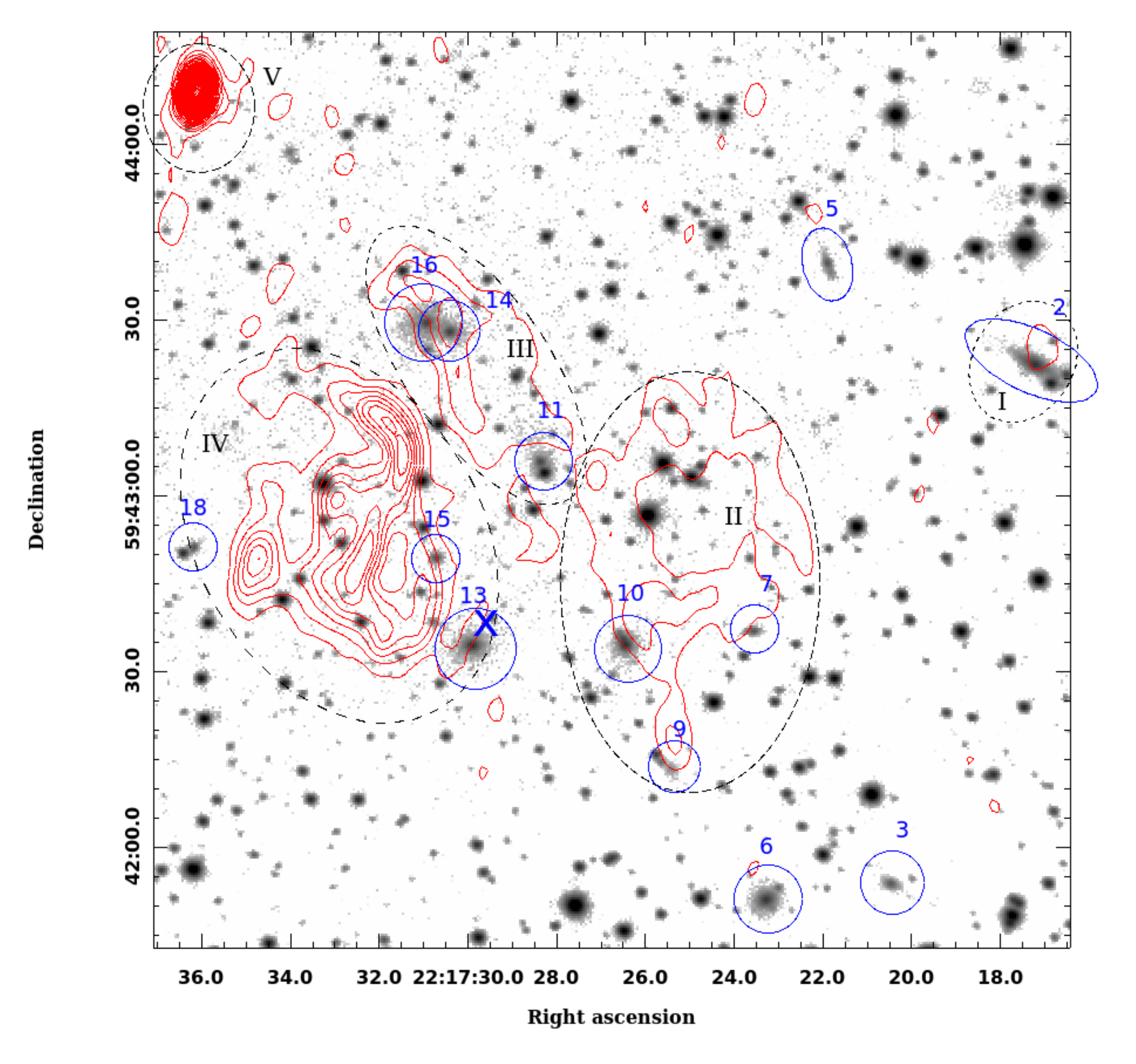}
\caption{The $2\farcm6 \times 2\farcm6$ field containing both the radio source 24P73 and the central part of the galaxy concentration with the brightest galaxies G2, G6, G13, and G16. Galaxies are indicated by their blue colour. The inverted grey-scale background image is the stack of the J-, H-, and K-band images from UGPS. 
The red curves are the contour lines of the GMRT 610 MHz radio map (lowest level at $3\sigma$, all others in steps of $5\sigma$ with $\sigma=55\,\rm \mu Jy$). Black dashed ellipses mark five radio components labelled with roman numbers. The luminosity-weighted centre of the projected galaxy distribution, except of G2, is indicated by the blue x symbol.
}
\label{fig:UGPS_Gray}
\end{figure*}
% 2016/images/crea_Grayscale_with_contours_and_grid_4.txt

\citet{Green_1994} presented a VLA L-band image at 1.5\,GHz where the radio source 24P73 is resolved into two distinct components, a brighter unresolved northern component and `faint, poorly defined emission extended over about an arcmin' that accounts for the bulk of the emission at 408 MHz.  Figure\,\ref{fig:UGPS_Gray} shows the part of the field from Fig.\,\ref{fig:UGPS_RGB} that contains both the radio emission and the brightest galaxies. For the sake of clarity, we distinguish five radio components that are marked by black ellipses and labelled I to V in Fig.\,\ref{fig:UGPS_Gray}.   The radio phoenix VLSS\,J2217.5+5943 described by \citet{vanWeeren_2009} and \citet{vanWeeren_2011} consists of the components II, III, and IV. Below, we briefly discuss the association of the radio components with galaxies. 
We start with the components II-IV, which are of central importance to the present study.

\subsection{Components II, III, and IV}

% Fig.09
\begin{figure}[htbp]
\includegraphics[width=0.48\textwidth]{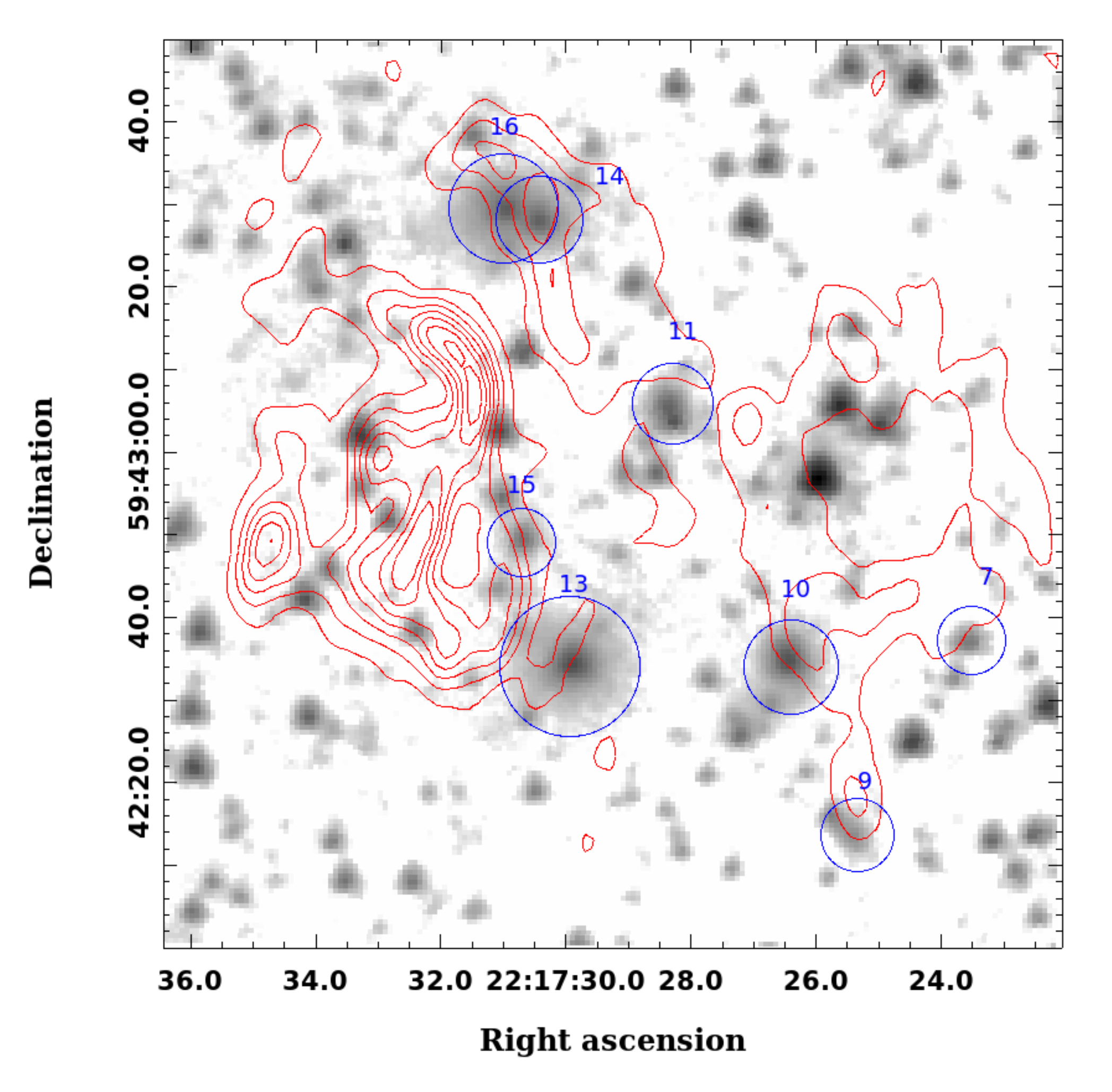}
\caption{610 MHz contours (as in Fig.\,\ref{fig:UGPS_Gray})  overplotted over the IRAC I1+I2 stacked image.  
}
\label{fig:IRAC_Gray}
\end{figure}

Our galaxy sample appears to be small. But this is again an expected consequence of the high foreground extinction in the innermost ZoA.
The number density of bright galaxies is clearly higher than the mean value \citep[][see below]{Frith_2003}.  
It can thus be assumed that our sample represents a cluster or group of galaxies. This interpretation is supported by the discovery of extended X-ray emission with Chandra (Mandal et al., submitted) as well as by the mass estimation (see Sect.\,\ref{sect:total_mass} below).

The majority of the galaxies in the sample, particularly the brighter ones, belong to an elongated structure that extends roughly from G1 in SW to G19 and G20 at the NE end.\footnote{With a length of $\sim 0.5$\,Mpc it is of roughly half the size of the chain of bright galaxies in the Perseus cluster. }
The brightest galaxies at either end are G6 and G16. 
Thus, it seems very likely that the concentration of galaxies seen in Fig.\,\ref{fig:UGPS_Gray} is connected with G6 and G16, rather than with G2. 
We deduce, therefore, that the redshift of the cluster is $z = 0.163 \pm .003$, corresponding to a luminosity distance of $D_{\rm L}$ = 752 Mpc and an angular scale of  2.7 kpc per arcsec.  The diameter of the region from which the galaxies were selected is thus $\sim 0.8$\,Mpc. 
Properties of the radio phoenix VLSS\,J2217.5+5943 derived with that redshift are given in Table\,\ref{tab:relic_prop}.

\begin{table}[htbp]
\caption{Observed properties of VLSS\,J2217.5+5943.}
\begin{tabular}{cccccc}
\hline\hline
\noalign{\smallskip}
         $z$               & $P_{1.4}$\tablefootmark{a}     &   LLS\tablefootmark{b}   &  $R_{\rm proj}$\tablefootmark{c}  & $\alpha_{\rm 38\,MHz}^{\rm 148\,MHz}$  & $\alpha_{\rm 32\,MHz}^{\rm 3\,GHz}$ \\
\hline
\noalign{\smallskip}
$0.163\pm.003$  & $1.0\pm .4$                                  &  270    &             68             &  1.2                               & 2.4 \\
\hline        
\vspace{-.7cm}
\end{tabular}
\tablefoot{
\tablefoottext{a}{Radio power in $10^{24}$ W\,Hz$^{-1}$ based on flux density $S_{\rm 1.4\,GHz}=12 \pm 3 \,\rm mJy$ from van Weeren et al. (2009) and a spectral slope of 2.4,    
see Fig.~\ref{fig:radio_spectrum}.}
\tablefoottext{b}{Largest linear size in kpc.}
\tablefoottext{c}{Projected distance from the cluster centre in kpc.}
}
\label{tab:relic_prop}
\end{table}

The X-ray emission does not show a significant peak and thus the dynamical state of the cluster is most likely unrelaxed (Mandal et al., submitted). 
Based on the distribution of the brightest galaxies, we classify the cluster as Rood-Sastry type L where three or more supergiant galaxies among the top 10 brightest galaxies are with comparable separation in a line. 
\citet{Wen_2013} calculated relaxation parameters of 2092 rich clusters and analysed statistical 
correlations with other cluster properties. 
They found that clusters of type L -- as well as types C, F, and I --  tend to be more unrelaxed than clusters with an outstandingly bright cD galaxy (type cD) or with two supergiant galaxies with a small separation (type B).

We used the K luminosities to compute the luminosity-weighted centre of the galaxies.
The K luminosity $L_{\rm K}$ of a galaxy is computed from its absolute magnitude $M_{\rm K} = K - DM - A_{\rm K} - k_{\rm K}$ adopting $M_{\rm K, \odot} = 3.27$ (Vega),  where $K$ is the measured Petrosion K magnitude, 
$DM = 5\log d_{\rm L}{\rm(Mpc)} + 25$ is the distance modulus, and $A_{\rm K}$  and  $k_{K}$ are  the interstellar extinction and the k-correction in the K band.  
Evolution corrections are neglected here. Adopting $A_K = 0.31\ E(B-V), \ E(B-V) = 1.80$ and the k corrections 
from \citet{Chilingarian_2012}\footnote{http://kcor.sai.msu.ru/}, 
we derived the absolute magnitudes given in Table\,\ref{tab:galaxies}.
The luminosity distance $d_{\rm L}$ was computed under the assumption $z = 0.163$ with the exception of  $z = 0.042$ for G2. 
The resulting luminosity-weighted centre at 
$22^{\rm h} 17^{\rm m} 29\fs5 +59\degr 42\arcmin 38\arcsec$ (J2000), indicated by the x symbol in Fig.~\ref{fig:UGPS_Gray}, is very close to both the bright galaxy G13 and the centre of the diffuse X-ray emission detected by Mandal et al. (submitted).  
According to Table\,\ref{tab:galaxies}, G13 and G6 are the two brightest cluster galaxies. 
Using the method described in Sect.\,\ref{sect:photo_redshift}, we derived a corrected magnitude $M_{\rm K} = -26.05$ for G13, 
very similar to the brightest cluster galaxy NGC\,1275 in the Perseus cluster. 

% Fig.10
\begin{figure}[htbp]
\includegraphics[width=9.2cm,angle=0]{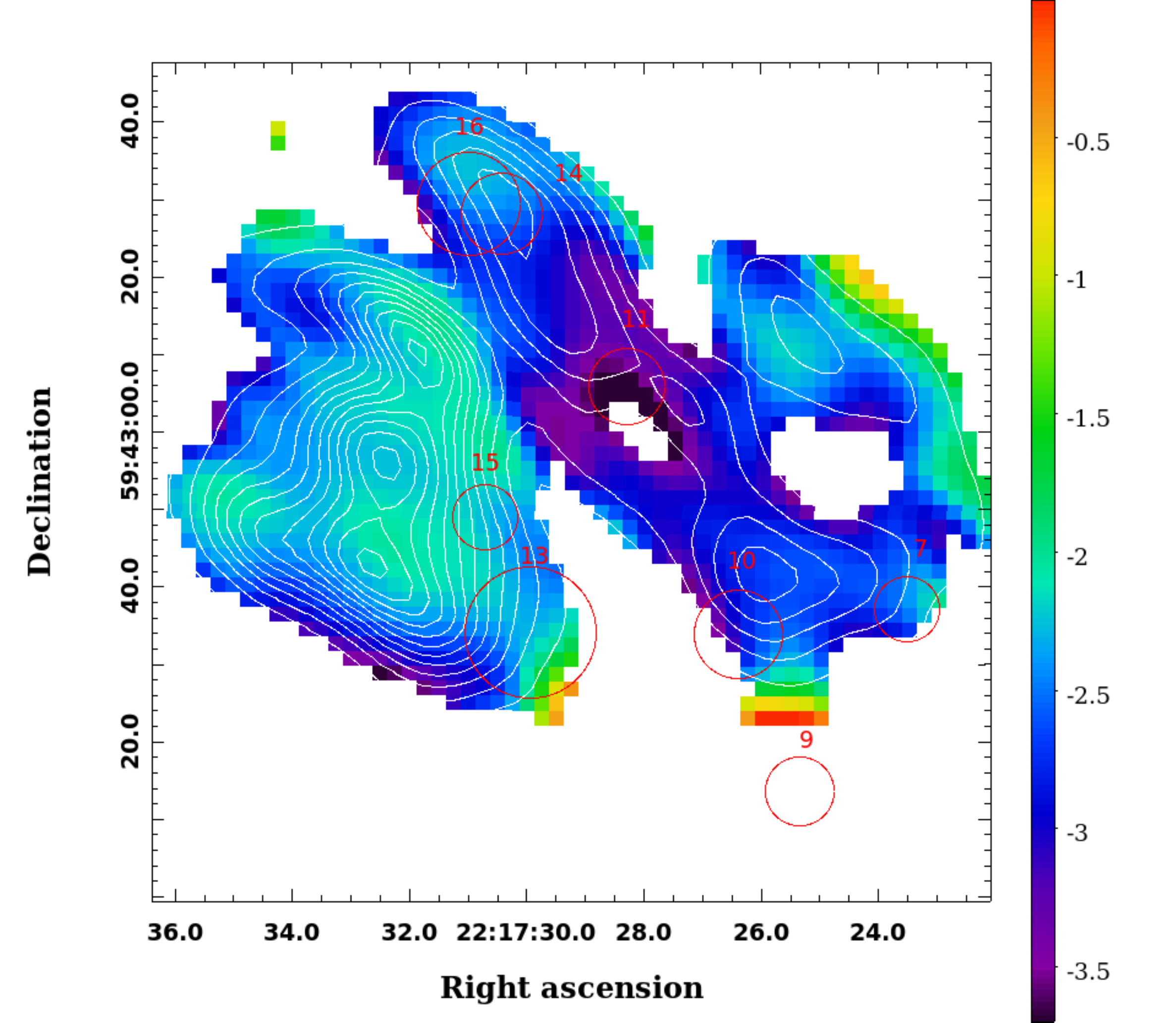}
\includegraphics[width=9.2cm,angle=0]{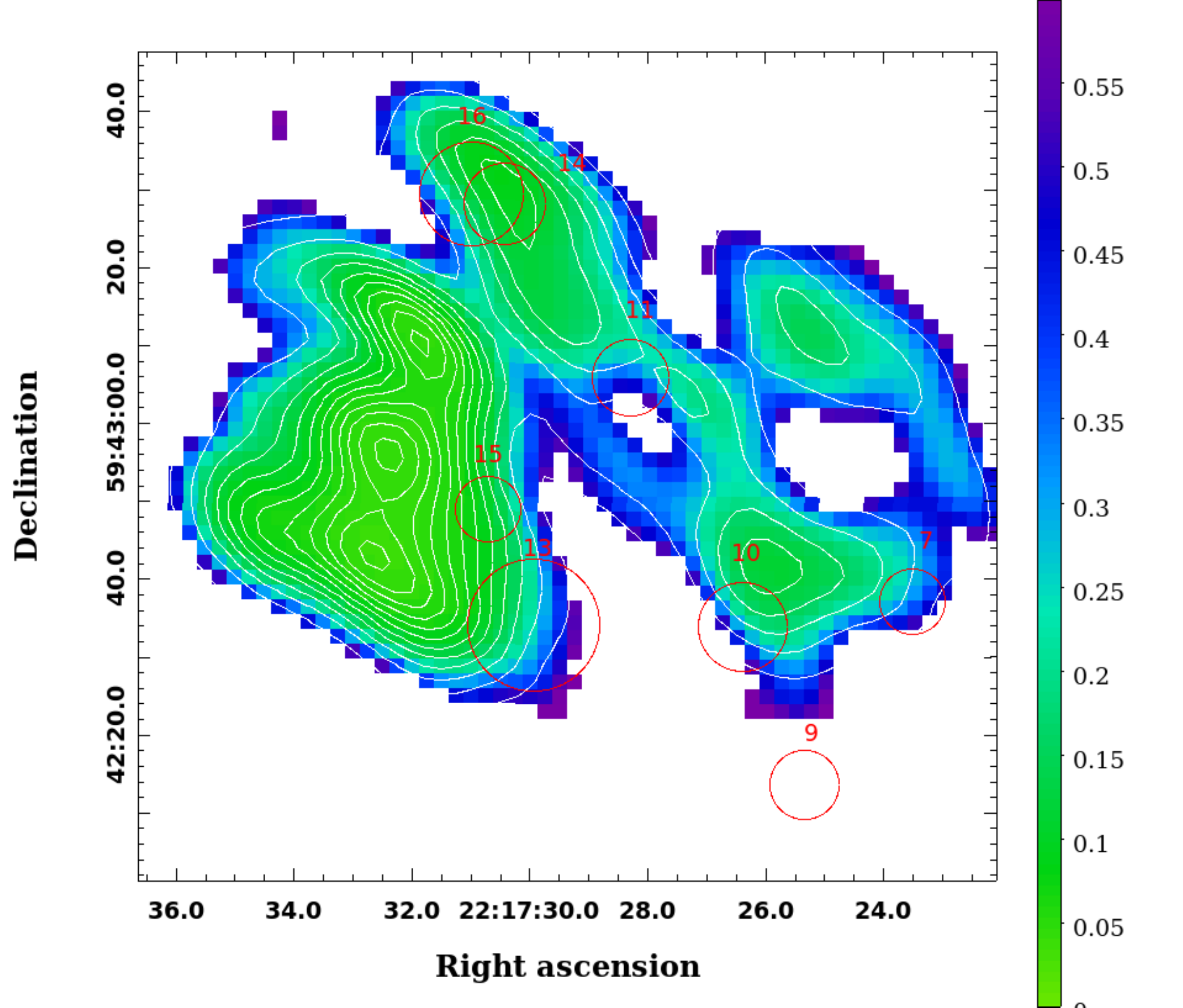}
\caption{Spectral index map (top)  and spectral  index error map (bottom) between 325 and 610 MHz with 325 MHz contours (white) .}
\label{fig:spectral_index_map}
\end{figure}

The extended emission region from  \citet{Green_1994} can be subdivided into the three different components II, III, and IV in the 610 MHz image, see Fig.~\ref{fig:UGPS_Gray}.
The largest angular scale of $\sim 100$\,arcsec, including the three components, corresponds to a (projected) largest linear size (LLS) of 270\,kpc. 
Component IV is by far the brightest part at 620\,MHz. It shows an irregular structure with several emission peaks. 
A separate faint peak is positionally coincident with the core of G13 and seems to indicate that this galaxy is active at some level. 
Remarkably, the other emission peaks are not associated with individual sources in the NIR (Fig.\,\ref{fig:UGPS_Gray}) or MIR  (Fig.\,\ref{fig:IRAC_Gray}). 
The closest projected distance of an emission peak from the centre of the galaxy distribution is $R_{\rm proj} = 68$\,kpc. The 1.4\,GHz radio power and the relatively small values for LLS and $R_{\rm proj}$ (Table\,\ref{tab:relic_prop}) are in good agreement with the corresponding values for the other radio phoenices in the \citet{vanWeeren_2009} sample.

The spectral index between 325 and 610 MHz shows little variation across  VLSS\,J2217.5+5943 (Fig.\,\ref{fig:spectral_index_map}). The radio spectrum is strongly bent, see Fig.\,\ref{fig:radio_spectrum} and also \citet{vanWeeren_2009}, indicating that the acceleration of electrons to relativistic energies has faded a significant amount of time ago. Assuming a field strength of about $5\,\rm \mu G$ in the emitting volume we estimate that the electron acceleration faded 60\,Myr ago. 
\citet{vanWeeren_2009} adopted an equipartition field strength of about $25\,\rm \mu G$ and estimated 30\,Myr. Conceivably, the radio feature VLSS\,J2217.5+5943 corresponds actually to remnant lobes of a former active phase of one of the galaxies in the cluster, see, e.g., \citet{2017A&A...606A..98B} for remnant radio galaxies from the LOFAR Lockman Hole field. However, the morphology of VLSS\,J2217.5+5943 is unusually distorted compared to other well studied dying radio galaxies in galaxy clusters \citep{2011A&A...526A.148M}. None of the galaxies we identified in the vicinity of VLSS\,J2217.5+5943 is a clear host galaxy, e.g. for symmetry reasons. Possibly, large velocities in the ICM have displaced and reshaped the plasma lobes emitted by one of the galaxies. See \citet{2021arXiv210204193V} for a current study of how ICM motions may advect the plasma lobes. The spectral ageing may have diminished the luminosity of the source at 1.4\,GHz significantly. If the source once had a radio power as expected from extrapolating from lower frequency, the luminosity was one order of magnitude higher than estimated in Tab.~\ref{tab:relic_prop}, still resulting in a luminosity very typical for radio galaxies. 

Component III shows a  nearly linear structure. The centre of the radio emission, approximately halfway along the linear extension, seems to be associated with G14. It is thus tempting to speculate that G14 is a  radio galaxy. Interestingly, G14 is very close to the bright galaxy G16 with a projected core distance of only 12 kpc at the redshift of the cluster. 
However, the spectral index map between 325 and 610 MHz, see Fig.~\ref{fig:spectral_index_map}, indicates that its spectrum is at least as steep as in component IV. Therefore,  we reject the idea that the radio emission of component III is predominantly associated with current AGN activity in this part. 

Component II appears to have an irregular structure of low surface brightness. With one exception, the subunits are not obviously associated with individual galaxies. The only exception is a faint and compact source at the southern border that seems to be positionally coincident with G9.

\subsection{Component I: the brightest galaxy G2}\label{sect:G2}

A faint radio source  is clearly visible close to the centre of the bright galaxy G2. 
Our optical spectrum of G2 allowed us to measure the diagnostic line ratios  [\ion{N}{ii}]\,6548,6584/H$\alpha$ and 
[\ion{S}{ii}]\,6717,6731/H$\alpha$. Based on the standard demarcation lines 
in the BPT diagrams \citep{Kewley_2001, Kauffmann_2003}, G2 is a LINER.
Moreover, there might be an additional weak broad H$\alpha$ component, indicating that it is a LINER type 1 galaxy.
G2 is the brightest and largest galaxy in the field. It can be obviously classified as a disk galaxy at a large inclination angle. 
The disk is traceable up to distance of 8-10 kpc from its centre.  
With a luminosity distance of 184 Mpc, compared to 742 Mpc for G6 and 762 Mpc for G16, G2 is obviously in the foreground. 
In addition, its projected position is distant from the centres of both the galaxy distribution (see below) and the diffuse X-ray emission.

\citet{Macri_2019} recently detected an isolated group or small cluster with redshift of $\sim 12,000$\,km\,s$^{-1}$  just outside of the innermost ZoA at $l \sim 100\degr$ and $b \sim -5\degr$.  This group of galaxies might be part of a larger structure when combining
them with G2 at  $l = 104\degr$ and $b = 2.4\degr$ with $ v = 12,600$\,km\,s$^{-1}$. 
The group from \citet{Macri_2019} is 25\,Mpc away from G2 and both have the same redshift. Nothing is known about galaxies in between. If they belong together and form a hidden supercluster, its diameter would be of the order of 25\,Mpc. This corresponds to the size of the Local Supercluster.

\subsection{Compact component V}\label{sect:compV}

In contrast to the other components, component V -- the compact radio source J221736.2+594407 -- has a typical AGN spectrum 
with $\alpha_{325}^{3000} = -0.86 \pm 0.05$. At lower frequencies, the spectrum is flatter with $\alpha_{150}^{325} \sim -0.4$  
\citep[see also][their Fig. 6]{vanWeeren_2011}, most likely due to synchrotron self-absorption.

\begin{figure}[htbp]
\includegraphics[width=9cm,angle=0]{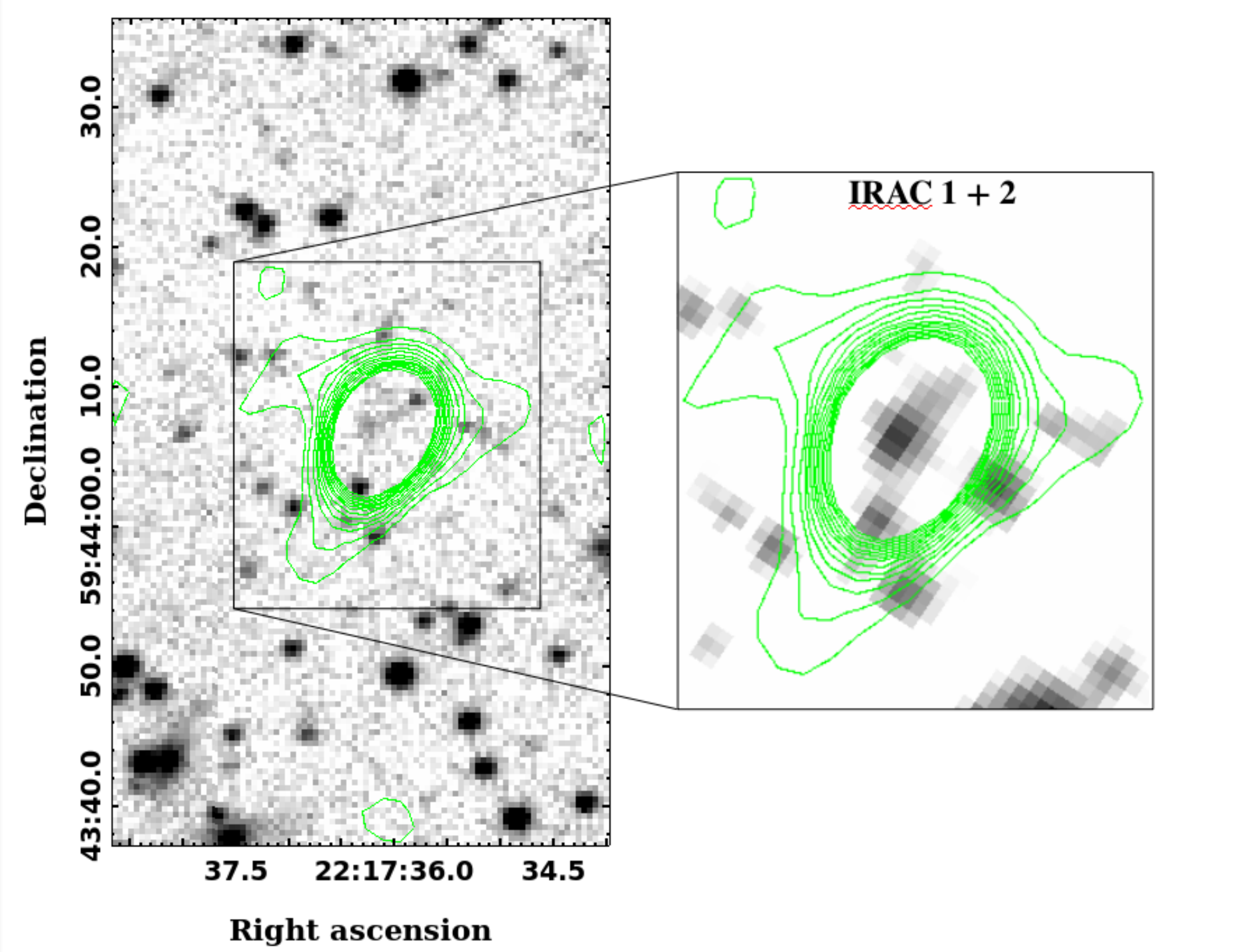}
\caption{The infrared counterpart of the compact radio source J221736.2+594407 (green contour lines) on the UGPS JHK image (left) and the IRAC 1 + 2 stack (right).  
Only the outer radio contours are plotted in order to show the NIR sources at the 610 MHz peak position (lowest contour at 3$\sigma$ local background, 
followed by steps of 5$\sigma$). 
}
\label{fig:Compact_source}
\end{figure}

On the UGPS image, a very faint NIR counterpart is seen, but not listed in the UKIDSSDR11PLUS database. 
At the same position, the Spitzer images in the IRAC 1 and 2 bands show a brighter source (Fig.\,\ref{fig:Compact_source}). The SMOG catalogue provided by the NASA/IRSA Infrared Science Archive (IRSA) lists the star-like source SSTSMOGC G104.6467+02.4411 at this position.
The source has the IRAC magnitudes  $[3.6] = 16.72 \pm 0.08$ and $[4.5] = 16.09 \pm 0.10$ (Vega system) but is too faint to be measured in the other two IRAC bands. 
Adopting $E(B-V) = 1.8$ and the extinction law measured in the Spitzer IRAC bands \citep{Indebetouw_2005, Chapman_2009},  
we find an extinction-corrected  colour index [3.6]-[4.5] (Vega) $= 0.55 \pm 0.17$, which is necessary but not sufficient for classifying the source as AGN  \citep{Stern_2005, Donley_2012}. 
We assume that the compact radio component at  J221736.2+594407 is most likely a background AGN that is not related to the galaxy cluster.

% Fig.11
% 2016/images/Greyscale_compact_radi0.odp

\subsection{Total stellar light, stellar mass, and cluster mass}\label{sect:total_mass}

The summed luminosity of all galaxies, except of G2, is $\sim 1.5\times10^{12}\, L_\odot$, where the unknown luminosity of G21 was simply set equal to that of G19. 
A more realistic estimate of the total luminosity of the cluster must take into account, in particular, 
the incompleteness of the galaxy selection and the possible contamination by foreground and background galaxies.  
We assume that the galaxy selection is complete for the brighter galaxies with $K \la 16$, corresponding to $L_{\rm K} \ga L_{\rm K, min}$. 
Then, we can estimate the total  stellar K-band luminosity by
\begin{equation}
  L_{\rm K} = f_{\rm c}\cdot (1 - f_{\rm bg})\cdot f_{\rm P}\cdot \sum_i L_{{\rm K}, i}(\ga L_{\rm K, min}), 
\end{equation}
where $L_{{\rm K}, i}(\ga L_{\rm K, min})$ is the luminosity of the galaxy $i$ that is brighter than $L_{\rm K, min}$.
The factor $f_{\rm c}$  includes the contribution from the galaxies fainter than $L_{\rm K, min}$, $f_{\rm bg}$ is the expected proportion of 
background galaxies\footnote{For simplicity, we refer to both foreground and background galaxies as background galaxies in the
following.}, and $f_{\rm P}$ corrects for  the fraction of light that is not covered by the Petrosian magnitudes of the galaxies.

The correction factor $f_{\rm c}$ can be estimated if the galaxy luminosity function (LF) $\Phi(L)$ is known. For the Schechter LF, 
 $\Phi(x) \propto x^{\, \alpha}\,e^{-x}$ with the two parameters $L^\ast$ and $\alpha$, it turns out that
 \begin{equation}
  f_{\rm c} = \Gamma(2+\alpha) \, \Bigg/ \int_{x_{\rm min}}^{\infty} x^{\, \alpha} e^{-x} dx,
 \end{equation}
where $\Gamma$ is the gamma function and $x = L/L^\ast$.  For $M_{\rm K, min} \approx -23.5$ and the parameters of the K-band LF for galaxies in low-z clusters 
from \citet{DePropris_2017} we find $f_{\rm c} \approx 2$. 
The surface density of background galaxies was taken from the number counts of 2MASS XSC galaxies presented by \citet{Frith_2003}.
Applied to our search field, we expect 6 galaxies brighter than the observed magnitude of $K \sim 16$, which corresponds to 
$f_{\rm bg} = 0.38$. 
In UKIDSS, the Petrosian flux is measured within a circular aperture to two times the Petrosian radius \citep{Dye_2006}.
The fraction of the galaxy's flux recovered by the Petrosian magnitudes amounts to $\sim 80$ \%  for the de Vaucouleurs profile \citep{Blanton_2001, Graham_2005}. 
We adopt a mean correction factor of $f_{\rm g} = 1.2 $.  
Therewith, the corrected total stellar luminosity amounts to  $L_{\rm K} \approx 3.3\times10^{12} L_\odot$ within $R = 400$\,kpc. 

It is generally accepted that the K-band luminosity is a good proxy for the stellar mass.
The mass in the stellar component of the cluster is given by  $\mathcal M_\ast = \langle {\mathcal M}/L_{\rm K}  \rangle  L_{\rm K, tot}$. 
The mean mass-to-luminosity ratio depends on the stellar populations in the galaxies. 
Here, we adopt $\langle {\mathcal M}/L_{\rm K}  \rangle = 0.8 {\mathcal M_\odot}/L_\odot$ 
from \citet{Graham_2013}, which results in $\mathcal M_\ast \approx 2.6\times10^{12} {\mathcal M_\odot}$.
The fraction of stars that were stripped from their host galaxies  and  feed the diffuse intra-cluster light was neglected.
The uncertainty of this estimation is difficult to grasp. 
Projection effects and the unknown cluster membership of the galaxies is the most problematic issue. 
Furthermore, the completeness correction $f_{\rm c}$ depends on the parameters of the LF, which may significantly differ from the mean cluster LF.

Observations over the past two decades have demonstrated that the stellar light and the stellar mass are useful proxies for the total halo mass 
$\mathcal{M}_{\rm h}$ of galaxy clusters or groups \citep[e.g.,][]{Lin_2003, Andreon_2012, Ziparo_2016, Wechsler_2018, Palmese_2020}. 
Based on the  $L_{\rm K} - {\mathcal M}_{\rm h}$ relation from  \citet{Ziparo_2016} we derive a total mass of  $1.5^{+0.7}_{-0.6}\times10^{14} {\mathcal M}_\odot$ within $R = 400$\,kpc, which corresponds to a high density contrast $\Delta = \rho(R)/\rho_{\rm c}(z) \approx 3.5\times10^3$, where $\rho_{\rm c}(z)$ is the critical mass density of the Universe at the redshift of the cluster.  Given that the stellar-to-total mass is radially constant on scales from about 150 kpc to 2.5 Mpc \citep{Andreon_2015} and adopting the NFW density profile, we roughly estimate 
${\mathcal M}_{\rm h, \Delta = 500} = 6.8^{+9.7}_{-4.8}\times10^{14}\,{\mathcal M}_\odot$ and $R_{500} = 1.4\pm 0.5$\,Mpc. 
The large errors reflect the uncertainties of the NFW profile parameters and the  $L_{\rm K} - {\mathcal M}_{\rm h}$ relation.
For comparison, the mean mass of the host clusters from the sample of 22 radio phoenices studied by Mandal et al. (submitted) 
is ${\mathcal M}_{\rm h, \Delta = 500} = (2.9\pm1.7)\times10^{14}\,{\mathcal M}_\odot$.
The radio relic is thus clearly within $R_{500}$, as was found for the other known phoenices \citep{vanWeeren_2009}.

%**********************************************************************************
%
\section{Summary and conclusions}
%
%**********************************************************************************

The radio source VLSS\,J2217.5+5943 (24P73) has been tentatively classified as a representative of the rare class of radio phoenices in galaxy clusters.  Because of its position in the innermost ZoA, a cluster has not been proven with certainty until now.
We exploited archival data from the UKIDSS Galactic Plane Survey in the NIR and from the Spitzer Mapping of the Outer Galaxy in the MIR
and selected a sample of 23 galaxies within a radius of 2.5 arcmin. 

We carried out spectroscopic observations of three of the brightest galaxies in the sample. For that aim we took advantage of the 
far-red/near-infrared integral field spectrograph LRS2-R at the 10\,m Hobby-Eberly Telescope, which is a superb instrument for securing spectra 
of faint red galaxies in the high-extinction regions close to the Galactic Plane. 
We were able to derive the redshift of $z = 0.165 \pm .001$ for the galaxy G16 that is spatially associated with 
VLSS\,J2217.5+5943. A second galaxy (G6) - just outside the radio emission region - is found to have a redshift of $z=0.161\pm .001$. 
The spectra and morphology of both systems correspond to E-type galaxies.
The third galaxy we observed (G2) -- at a distance of 120\,arcsec from G16 -- has turned out to be a foreground galaxy with
redshift of $z = 0.042$.  It is of LINER 1 type.  
All three galaxies are strongly reddened with a visual extinction of nearly 6\,mag due to Galactic dust in the innermost ZoA.

The galaxies G6 and G16 belong to a chain of more than 10 galaxies, among them 8 galaxies brighter than $M_{\rm K} = -24$.
We suggest that the chain represents a galaxy cluster at $z = 0.163\pm .003$. 
A similar redshift was anticipated by \citet{vanWeeren_2011},  but with a significant uncertainty.
The cluster mass is roughly estimated at ${\mathcal M}_{\rm h, \Delta = 500} = 6.8^{+9.7}_{-4.8}\,10^{14}\,{\mathcal M}_\odot$. 
The radio source VLSS\,J2217.5+5943 borders on the main part of the galaxy chain. 
The presence of a galaxy cluster, as indicated by our observations, corroborates that the source VLSS\,J2217.5+5943 became distorted and possibly enhanced in luminosity by the ICM.  The largest projected linear extension of the radio phoenix is $\sim 270$\,kpc. 
The projected distance from the assumed cluster centre to the closest radio emission peak is 68\,kpc.  
These properties are very similar to those of the other known radio phoenices.  
According to its spectrum and luminosity the source could be a remnant radio galaxy, severely distorted by ICM motions possibly induced by a cluster merger. The presence of a cluster would explain the complex morphology of the source. Our observations provide evidence that this is actually the case.

\begin{acknowledgements}

This work has been supported by the DFG grants KO857/33-1 and CH71/34-3.
This paper is based on observations obtained with the Hobby-Eberly Telescope, which is a joint project of the University of Texas at Austin, the Pennsylvania State University, Ludwig-Maximilians-Universität M\"unchen, and Georg-August-Universit\"at G\"ottingen.
The Low Resolution Spectrograph 2 (LRS2) was developed and funded by the University of Texas at Austin McDonald Observatory and Department of Astronomy and by Pennsylvania State University. We thank the Leibniz-Institut für Astrophysik Potsdam (AIP) and the Institut für Astrophysik Göttingen (IAG) for their contributions to the construction of the integral field units.
This  paper  makes  extensive  use  of  the  United Kingdom Infrared Telescope (UKIRT)  Infrared Deep Sky Survey (UKIDSS).
UKIRT is operated by the Joint Astronomy Centre on behalf of the Science and Technology Facilities Council of the U.K.  
This research made use of Montage. It is funded by
the National Science Foundation under Grant Number ACI-1440620, and was previously
funded by the National Aeronautics and Space Administration's Earth Science
Technology Office, Computation Technologies Project, under Cooperative Agreement
Number NCC5-626 between NASA and the California Institute of Technology.

This publication has also made use of data products from
the VizieR catalogue access tool, CDS, Strasbourg, France and of data obtained from 
the NASA/IPAC Infrared Science Archive (IRSA), operated by the 
Jet Propulsion Laboratories/California Institute of Technology, founded
by the National Aeronautic and Space Administration. In particular,  
This work is based in part on observations made with the Spitzer Space Telescope, 
which is operated by the Jet Propulsion Laboratory, California Institute of Technology under a contract with NASA. 

In addition, data from  the Pan-STARRS1 Surveys (PS1) were used in this paper.
The Pan-STARRS1 Surveys (PS1) and the PS1 public science archive have been made possible 
through contributions by the Institute for Astronomy, the University of Hawaii, the Pan-STARRS Project Office, 
the Max-Planck Society and its participating institutes, the Max Planck Institute for Astronomy, Heidelberg 
and the Max Planck Institute for Extraterrestrial Physics, Garching, The Johns Hopkins University, 
Durham University, the University of Edinburgh, the Queen's University Belfast, the Harvard-Smithsonian Center for Astrophysics, 
the Las Cumbres Observatory Global Telescope Network Incorporated, the National Central University of Taiwan, 
the Space Telescope Science Institute, the National Aeronautics and Space Administration 
under Grant No. NNX08AR22G issued through the Planetary Science Division of the 
NASA Science Mission Directorate, the National Science Foundation Grant No. AST-1238877, 
the University of Maryland, Eotvos Lorand University (ELTE), the Los Alamos National Laboratory, and the Gordon and Betty Moore Foundation.

\end{acknowledgements}

%% WARNING
%%-------------------------------------------------------------------
%% Please note that we have included the references to the file aa.dem in
%% order to compile it, but we ask you to:
%%
%% - use BibTeX with the regular commands:
%%   \bibliographystyle{aa} % style aa.bst
%%   \bibliography{Yourfile} % your references Yourfile.bib
%%
%% - join the .bib files when you upload your source files
%-------------------------------------------------------------------

\bibliographystyle{aa} % style aa.bst
\bibliography{literature} % your references Yourfile.bib

\begin{thebibliography}{68}
\expandafter\ifx\csname natexlab\endcsname\relax\def\natexlab#1{#1}\fi

\bibitem[{{Andreon}(2012)}]{Andreon_2012}
{Andreon}, S. 2012, \aap, 548, A83

\bibitem[{{Andreon}(2015)}]{Andreon_2015}
{Andreon}, S. 2015, \aap, 575, A108

\bibitem[{{Barentsen} {et~al.}(2014){Barentsen}, {Farnhill}, {Drew},
  {Gonz{\'a}lez-Solares}, {Greimel}, {Irwin}, {Miszalski}, {Ruhland}, {Groot},
  {Mampaso}, {Sale}, {Henden}, {Aungwerojwit}, {Barlow}, {Carter}, {Corradi},
  {Drake}, {Eisl{\"o}ffel}, {Fabregat}, {G{\"a}nsicke}, {Gentile Fusillo},
  {Greiss}, {Hales}, {Hodgkin}, {Huckvale}, {Irwin}, {King}, {Knigge},
  {Kupfer}, {Lagadec}, {Lennon}, {Lewis}, {Mohr-Smith}, {Morris}, {Naylor},
  {Parker}, {Phillipps}, {Pyrzas}, {Raddi}, {Roelofs}, {Rodr{\'\i}guez-Gil},
  {Sabin}, {Scaringi}, {Steeghs}, {Suso}, {Tata}, {Unruh}, {van Roestel},
  {Viironen}, {Vink}, {Walton}, {Wright}, \& {Zijlstra}}]{Barentsen_2014}
{Barentsen}, G., {Farnhill}, H.~J., {Drew}, J.~E., {et~al.} 2014, \mnras, 444,
  3230

\bibitem[{{Bessell} {et~al.}(1998){Bessell}, {Castelli}, \&
  {Plez}}]{1998A&A...333..231B}
{Bessell}, M.~S., {Castelli}, F., \& {Plez}, B. 1998, \aap, 333, 231

\bibitem[{{Blanton} {et~al.}(2001){Blanton}, {Dalcanton}, {Eisenstein},
  {Loveday}, {Strauss}, {SubbaRao}, {Weinberg}, {Anderson}, {Annis}, {Bahcall},
  {Bernardi}, {Brinkmann}, {Brunner}, {Burles}, {Carey}, {Castander},
  {Connolly}, {Csabai}, {Doi}, {Finkbeiner}, {Friedman}, {Frieman}, {Fukugita},
  {Gunn}, {Hennessy}, {Hindsley}, {Hogg}, {Ichikawa}, {Ivezi{\'c}}, {Kent},
  {Knapp}, {Lamb}, {Leger}, {Long}, {Lupton}, {McKay}, {Meiksin}, {Merelli},
  {Munn}, {Narayanan}, {Newcomb}, {Nichol}, {Okamura}, {Owen}, {Pier}, {Pope},
  {Postman}, {Quinn}, {Rockosi}, {Schlegel}, {Schneider}, {Shimasaku},
  {Siegmund}, {Smee}, {Snir}, {Stoughton}, {Stubbs}, {Szalay}, {Szokoly},
  {Thakar}, {Tremonti}, {Tucker}, {Uomoto}, {Vanden Berk}, {Vogeley},
  {Waddell}, {Yanny}, {Yasuda}, \& {York}}]{Blanton_2001}
{Blanton}, M.~R., {Dalcanton}, J., {Eisenstein}, D., {et~al.} 2001, \aj, 121,
  2358

\bibitem[{{Brienza} {et~al.}(2017){Brienza}, {Godfrey}, {Morganti}, {Prandoni},
  {Harwood}, {Mahony}, {Hardcastle}, {Murgia}, {R{\"o}ttgering}, {Shimwell}, \&
  {Shulevski}}]{2017A&A...606A..98B}
{Brienza}, M., {Godfrey}, L., {Morganti}, R., {et~al.} 2017, \aap, 606, A98

\bibitem[{{Carey} {et~al.}(2008){Carey}, {Ali}, {Berriman}, {Boulanger},
  {Brunt}, {Cutri}, {Flagey}, {Gibson}, {Heyer}, {Hora}, {Indebetouw},
  {Kraemer}, {Kuchar}, {Latter}, {Marleau}, {Miville-Deschenes}, {Mizuno},
  {Molinari}, {Noriega-Crespo}, {Padgett}, {Paladini}, {Price}, {Rebull},
  {Rottler}, {Shenoy}, {Shipman}, \& {Testi}}]{Carey_2008}
{Carey}, S., {Ali}, B., {Berriman}, B., {et~al.} 2008, {Spitzer Mapping of the
  Outer Galaxy (SMOG)}, Spitzer Proposal

\bibitem[{{Casali} {et~al.}(2007){Casali}, {Adamson}, {Alves de Oliveira},
  {Almaini}, {Burch}, {Chuter}, {Elliot}, {Folger}, {Foucaud}, {Hambly},
  {Hastie}, {Henry}, {Hirst}, {Irwin}, {Ives}, {Lawrence}, {Laidlaw}, {Lee},
  {Lewis}, {Lunney}, {McLay}, {Montgomery}, {Pickup}, {Read}, {Rees}, {Robson},
  {Sekiguchi}, {Vick}, {Warren}, \& {Woodward}}]{Casali_2007}
{Casali}, M., {Adamson}, A., {Alves de Oliveira}, C., {et~al.} 2007, \aap, 467,
  777

\bibitem[{{Chambers} {et~al.}(2019){Chambers}, {Magnier}, {Metcalfe},
  {Flewelling}, {Huber}, {Waters}, {Denneau}, {Draper}, {Farrow}, {Finkbeiner},
  {Holmberg}, {Koppenhoefer}, {Price}, {Rest}, {Saglia}, {Schlafly}, {Smartt},
  {Sweeney}, {Wainscoat}, {Burgett}, {Chastel}, {Grav}, {Heasley}, {Hodapp},
  {Jedicke}, {Kaiser}, {Kudritzki}, {Luppino}, {Lupton}, {Monet}, {Morgan},
  {Onaka}, {Shiao}, {Stubbs}, {Tonry}, {White}, {Ba{\~n}ados}, {Bell},
  {Bender}, {Bernard}, {Boegner}, {Boffi}, {Botticella}, {Calamida},
  {Casertano}, {Chen}, {Chen}, {Cole}, {Deacon}, {Frenk}, {Fitzsimmons},
  {Gezari}, {Gibbs}, {Goessl}, {Goggia}, {Gourgue}, {Goldman}, {Grant},
  {Grebel}, {Hambly}, {Hasinger}, {Heavens}, {Heckman}, {Henderson}, {Henning},
  {Holman}, {Hopp}, {Ip}, {Isani}, {Jackson}, {Keyes}, {Koekemoer}, {Kotak},
  {Le}, {Liska}, {Long}, {Lucey}, {Liu}, {Martin}, {Masci}, {McLean}, {Mindel},
  {Misra}, {Morganson}, {Murphy}, {Obaika}, {Narayan}, {Nieto-Santisteban},
  {Norberg}, {Peacock}, {Pier}, {Postman}, {Primak}, {Rae}, {Rai}, {Riess},
  {Riffeser}, {Rix}, {R{\"o}ser}, {Russel}, {Rutz}, {Schilbach}, {Schultz},
  {Scolnic}, {Strolger}, {Szalay}, {Seitz}, {Small}, {Smith}, {Soderblom},
  {Taylor}, {Thomson}, {Taylor}, {Thakar}, {Thiel}, {Thilker}, {Unger},
  {Urata}, {Valenti}, {Wagner}, {Walder}, {Walter}, {Watters}, {Werner},
  {Wood-Vasey}, \& {Wyse}}]{Chambers_2019}
{Chambers}, K.~C., {Magnier}, E.~A., {Metcalfe}, N., {et~al.} 2019, arXiv
  e-prints, arXiv:1612.05560v4

\bibitem[{{Chapman} {et~al.}(2009){Chapman}, {Mundy}, {Lai}, \&
  {Evans}}]{Chapman_2009}
{Chapman}, N.~L., {Mundy}, L.~G., {Lai}, S.-P., \& {Evans}, Neal~J., I. 2009,
  \apj, 690, 496

\bibitem[{{Chilingarian} \& {Zolotukhin}(2012)}]{Chilingarian_2012}
{Chilingarian}, I.~V. \& {Zolotukhin}, I.~Y. 2012, \mnras, 419, 1727

\bibitem[{{Chonis} {et~al.}(2016){Chonis}, {Hill}, {Lee}, {Tuttle}, {Vattiat},
  {Drory}, {Indahl}, {Peterson}, \& {Ramsey}}]{chonis_2016}
{Chonis}, T.~S., {Hill}, G.~J., {Lee}, H., {et~al.} 2016, Society of
  Photo-Optical Instrumentation Engineers (SPIE) Conference Series, Vol. 9908,
  {LRS2: design, assembly, testing, and commissioning of the second-generation
  low-resolution spectrograph for the Hobby-Eberly Telescope}, 99084C

\bibitem[{{De Propris}(2017)}]{DePropris_2017}
{De Propris}, R. 2017, \mnras, 465, 4035

\bibitem[{{Donley} {et~al.}(2012){Donley}, {Koekemoer}, {Brusa}, {Capak},
  {Cardamone}, {Civano}, {Ilbert}, {Impey}, {Kartaltepe}, {Miyaji}, {Salvato},
  {Sanders}, {Trump}, \& {Zamorani}}]{Donley_2012}
{Donley}, J.~L., {Koekemoer}, A.~M., {Brusa}, M., {et~al.} 2012, \apj, 748, 142

\bibitem[{{Dressler} {et~al.}(1987){Dressler}, {Lynden-Bell}, {Burstein},
  {Davies}, {Faber}, {Terlevich}, \& {Wegner}}]{Dressler_1987}
{Dressler}, A., {Lynden-Bell}, D., {Burstein}, D., {et~al.} 1987, \apj, 313, 42

\bibitem[{{Dye} {et~al.}(2006){Dye}, {Warren}, {Hambly}, {Cross}, {Hodgkin},
  {Irwin}, {Lawrence}, {Adamson}, {Almaini}, {Edge}, {Hirst}, {Jameson},
  {Lucas}, {van Breukelen}, {Bryant}, {Casali}, {Collins}, {Dalton}, {Davies},
  {Davis}, {Emerson}, {Evans}, {Foucaud}, {Gonzales-Solares}, {Hewett},
  {Kendall}, {Kerr}, {Leggett}, {Lodieu}, {Loveday}, {Lewis}, {Mann},
  {McMahon}, {Mortlock}, {Nakajima}, {Pinfield}, {Rawlings}, {Read}, {Riello},
  {Sekiguchi}, {Smith}, {Sutorius}, {Varricatt}, {Walton}, \&
  {Weatherley}}]{Dye_2006}
{Dye}, S., {Warren}, S.~J., {Hambly}, N.~C., {et~al.} 2006, \mnras, 372, 1227

\bibitem[{{Fazio} {et~al.}(2004){Fazio}, {Hora}, {Allen}, {Ashby}, {Barmby},
  {Deutsch}, {Huang}, {Kleiner}, {Marengo}, {Megeath}, {Melnick}, {Pahre},
  {Patten}, {Polizotti}, {Smith}, {Taylor}, {Wang}, {Willner}, {Hoffmann},
  {Pipher}, {Forrest}, {McMurty}, {McCreight}, {McKelvey}, {McMurray}, {Koch},
  {Moseley}, {Arendt}, {Mentzell}, {Marx}, {Losch}, {Mayman}, {Eichhorn},
  {Krebs}, {Jhabvala}, {Gezari}, {Fixsen}, {Flores}, {Shakoorzadeh}, {Jungo},
  {Hakun}, {Workman}, {Karpati}, {Kichak}, {Whitley}, {Mann}, {Tollestrup},
  {Eisenhardt}, {Stern}, {Gorjian}, {Bhattacharya}, {Carey}, {Nelson},
  {Glaccum}, {Lacy}, {Lowrance}, {Laine}, {Reach}, {Stauffer}, {Surace},
  {Wilson}, {Wright}, {Hoffman}, {Domingo}, \& {Cohen}}]{Fazio_2004}
{Fazio}, G.~G., {Hora}, J.~L., {Allen}, L.~E., {et~al.} 2004, \apjs, 154, 10

\bibitem[{{Flewelling} {et~al.}(2020){Flewelling}, {Magnier}, {Chambers},
  {Heasley}, {Holmberg}, {Huber}, {Sweeney}, {Waters}, {Calamida}, {Casertano},
  {Chen}, {Farrow}, {Hasinger}, {Henderson}, {Long}, {Metcalfe}, {Narayan},
  {Nieto-Santisteban}, {Norberg}, {Rest}, {Saglia}, {Szalay}, {Thakar},
  {Tonry}, {Valenti}, {Werner}, {White}, {Denneau}, {Draper}, {Hodapp},
  {Jedicke}, {Kaiser}, {Kudritzki}, {Price}, {Wainscoat}, {Chastel}, {McLean},
  {Postman}, \& {Shiao}}]{Flewelling_2020}
{Flewelling}, H.~A., {Magnier}, E.~A., {Chambers}, K.~C., {et~al.} 2020, \apjs,
  251, 7

\bibitem[{{Frith} {et~al.}(2003){Frith}, {Busswell}, {Fong}, {Metcalfe}, \&
  {Shanks}}]{Frith_2003}
{Frith}, W.~J., {Busswell}, G.~S., {Fong}, R., {Metcalfe}, N., \& {Shanks}, T.
  2003, \mnras, 345, 1049

\bibitem[{{Giovanelli} \& {Haynes}(1982)}]{Giovanelli_1982}
{Giovanelli}, R. \& {Haynes}, M.~P. 1982, \aj, 87, 1355

\bibitem[{{Graham} \& {Driver}(2005)}]{Graham_2005}
{Graham}, A.~W. \& {Driver}, S.~P. 2005, \pasa, 22, 118

\bibitem[{{Graham} \& {Scott}(2013)}]{Graham_2013}
{Graham}, A.~W. \& {Scott}, N. 2013, \apj, 764, 151

\bibitem[{{Green} \& {Joncas}(1994)}]{Green_1994}
{Green}, D.~A. \& {Joncas}, G. 1994, \aaps, 104, 481

\bibitem[{{Green} {et~al.}(2004){Green}, {Lacy}, {Bhatnagar}, {Gates}, \&
  {Warner}}]{Green_2004}
{Green}, D.~A., {Lacy}, M., {Bhatnagar}, S., {Gates}, E.~L., \& {Warner}, P.~J.
  2004, \mnras, 354, 1159

\bibitem[{{Hales} {et~al.}(1995){Hales}, {Waldram}, {Rees}, \&
  {Warner}}]{1995MNRAS.274..447H}
{Hales}, S.~E.~G., {Waldram}, E.~M., {Rees}, N., \& {Warner}, P.~J. 1995,
  \mnras, 274, 447

\bibitem[{{Hambly} {et~al.}(2008){Hambly}, {Collins}, {Cross}, {Mann}, {Read},
  {Sutorius}, {Bond}, {Bryant}, {Emerson}, {Lawrence}, {Rimoldini}, {Stewart},
  {Williams}, {Adamson}, {Hirst}, {Dye}, \& {Warren}}]{Hambly_2008}
{Hambly}, N.~C., {Collins}, R.~S., {Cross}, N.~J.~G., {et~al.} 2008, \mnras,
  384, 637

\bibitem[{{Hewett} {et~al.}(2006){Hewett}, {Warren}, {Leggett}, \&
  {Hodgkin}}]{Hewett_2006}
{Hewett}, P.~C., {Warren}, S.~J., {Leggett}, S.~K., \& {Hodgkin}, S.~T. 2006,
  \mnras, 367, 454

\bibitem[{{Higgs} \& {Joncas}(1989)}]{Higgs_1989}
{Higgs}, L.~A. \& {Joncas}, G. 1989, \jrasc, 83, 313

\bibitem[{{Hill} {et~al.}(2018){Hill}, {Drory}, {Good}, {Lee}, {Vattiat},
  {Kriel}, {Ramsey}, {Bryant}, {Fowler}, {Landriau}, {Leck}, {Mrozinski},
  {Odewahn}, {Shetrone}, {Westfall}, {Terrazas}, {Balderrama}, {Bevins},
  {Buetow}, {Caldwell}, {Damm}, {MacQueen}, {Martin}, {Martin}, {Pautzke},
  {Smither}, {Rostopchin}, {Smith}, {Spencer}, {Armandroff}, {Gebhardt}, \&
  {Ramsey}}]{hill_2018}
{Hill}, G.~J., {Drory}, N., {Good}, J.~M., {et~al.} 2018, in Society of
  Photo-Optical Instrumentation Engineers (SPIE) Conference Series, Vol. 10700,
  \procspie, 107000P

\bibitem[{{Indebetouw} {et~al.}(2005){Indebetouw}, {Mathis}, {Babler}, {Meade},
  {Watson}, {Whitney}, {Wolff}, {Wolfire}, {Cohen}, {Bania}, {Benjamin},
  {Clemens}, {Dickey}, {Jackson}, {Kobulnicky}, {Marston}, {Mercer},
  {Stauffer}, {Stolovy}, \& {Churchwell}}]{Indebetouw_2005}
{Indebetouw}, R., {Mathis}, J.~S., {Babler}, B.~L., {et~al.} 2005, \apj, 619,
  931

\bibitem[{{Intema} {et~al.}(2017){Intema}, {Jagannathan}, {Mooley}, \&
  {Frail}}]{2017A&A...598A..78I}
{Intema}, H.~T., {Jagannathan}, P., {Mooley}, K.~P., \& {Frail}, D.~A. 2017,
  \aap, 598, A78

\bibitem[{{Intema} {et~al.}(2009){Intema}, {van der Tol}, {Cotton}, {Cohen},
  {van Bemmel}, \& {R{\"o}ttgering}}]{2009A&A...501.1185I}
{Intema}, H.~T., {van der Tol}, S., {Cotton}, W.~D., {et~al.} 2009, \aap, 501,
  1185

\bibitem[{{Kauffmann} {et~al.}(2003){Kauffmann}, {Heckman}, {Tremonti},
  {Brinchmann}, {Charlot}, {White}, {Ridgway}, {Brinkmann}, {Fukugita}, {Hall},
  {Ivezi{\'c}}, {Richards}, \& {Schneider}}]{Kauffmann_2003}
{Kauffmann}, G., {Heckman}, T.~M., {Tremonti}, C., {et~al.} 2003, \mnras, 346,
  1055

\bibitem[{{Kewley} {et~al.}(2001){Kewley}, {Dopita}, {Sutherland}, {Heisler},
  \& {Trevena}}]{Kewley_2001}
{Kewley}, L.~J., {Dopita}, M.~A., {Sutherland}, R.~S., {Heisler}, C.~A., \&
  {Trevena}, J. 2001, \apj, 556, 121

\bibitem[{{Kinney} {et~al.}(1996){Kinney}, {Calzetti}, {Bohlin}, {McQuade},
  {Storchi-Bergmann}, \& {Schmitt}}]{Kinney_1996}
{Kinney}, A.~L., {Calzetti}, D., {Bohlin}, R.~C., {et~al.} 1996, \apj, 467, 38

\bibitem[{{Kirkpatrick} {et~al.}(1991){Kirkpatrick}, {Henry}, \&
  {McCarthy}}]{Kirkpatrick_1991}
{Kirkpatrick}, J.~D., {Henry}, T.~J., \& {McCarthy}, Donald~W., J. 1991, \apjs,
  77, 417

\bibitem[{{Kocevski} {et~al.}(2007){Kocevski}, {Ebeling}, {Mullis}, \&
  {Tully}}]{Kocevski_2007}
{Kocevski}, D.~D., {Ebeling}, H., {Mullis}, C.~R., \& {Tully}, R.~B. 2007,
  \apj, 662, 224

\bibitem[{{Kraan-Korteweg}(2005)}]{Kraan_2005}
{Kraan-Korteweg}, R.~C. 2005, Reviews in Modern Astronomy, 18, 48

\bibitem[{{Kraan-Korteweg} {et~al.}(2018){Kraan-Korteweg}, {van Driel},
  {Schr{\"o}der}, {Ramatsoku}, \& {Henning}}]{Kraan_2018}
{Kraan-Korteweg}, R.~C., {van Driel}, W., {Schr{\"o}der}, A.~C., {Ramatsoku},
  M., \& {Henning}, P.~A. 2018, \mnras, 481, 1262

\bibitem[{{Lacy} {et~al.}(2020){Lacy}, {Baum}, {Chandler}, {Chatterjee},
  {Clarke}, {Deustua}, {English}, {Farnes}, {Gaensler}, {Gugliucci},
  {Hallinan}, {Kent}, {Kimball}, {Law}, {Lazio}, {Marvil}, {Mao}, {Medlin},
  {Mooley}, {Murphy}, {Myers}, {Osten}, {Richards}, {Rosolowsky}, {Rudnick},
  {Schinzel}, {Sivakoff}, {Sjouwerman}, {Taylor}, {White}, {Wrobel},
  {Andernach}, {Beasley}, {Berger}, {Bhatnager}, {Birkinshaw}, {Bower},
  {Brandt}, {Brown}, {Burke-Spolaor}, {Butler}, {Comerford}, {Demorest}, {Fu},
  {Giacintucci}, {Golap}, {G{\"u}th}, {Hales}, {Hiriart}, {Hodge}, {Horesh},
  {Ivezi{\'c}}, {Jarvis}, {Kamble}, {Kassim}, {Liu}, {Loinard}, {Lyons},
  {Masters}, {Mezcua}, {Moellenbrock}, {Mroczkowski}, {Nyland}, {O'Dea},
  {O'Sullivan}, {Peters}, {Radford}, {Rao}, {Robnett}, {Salcido}, {Shen},
  {Sobotka}, {Witz}, {Vaccari}, {van Weeren}, {Vargas}, {Williams}, \&
  {Yoon}}]{2020PASP..132c5001L}
{Lacy}, M., {Baum}, S.~A., {Chandler}, C.~J., {et~al.} 2020, \pasp, 132, 035001

\bibitem[{{Lambert} {et~al.}(2020){Lambert}, {Kraan-Korteweg}, {Jarrett}, \&
  {Macri}}]{Lambert_2020}
{Lambert}, T.~S., {Kraan-Korteweg}, R.~C., {Jarrett}, T.~H., \& {Macri}, L.~M.
  2020, \mnras, 497, 2954

\bibitem[{{Lane} {et~al.}(2014){Lane}, {Cotton}, {van Velzen}, {Clarke},
  {Kassim}, {Helmboldt}, {Lazio}, \& {Cohen}}]{2014MNRAS.440..327L}
{Lane}, W.~M., {Cotton}, W.~D., {van Velzen}, S., {et~al.} 2014, \mnras, 440,
  327

\bibitem[{{Lawrence} {et~al.}(2007){Lawrence}, {Warren}, {Almaini}, {Edge},
  {Hambly}, {Jameson}, {Lucas}, {Casali}, {Adamson}, {Dye}, {Emerson},
  {Foucaud}, {Hewett}, {Hirst}, {Hodgkin}, {Irwin}, {Lodieu}, {McMahon},
  {Simpson}, {Smail}, {Mortlock}, \& {Folger}}]{Lawrence_2007}
{Lawrence}, A., {Warren}, S.~J., {Almaini}, O., {et~al.} 2007, \mnras, 379,
  1599

\bibitem[{{Lin} {et~al.}(2003){Lin}, {Mohr}, \& {Stanford}}]{Lin_2003}
{Lin}, Y.-T., {Mohr}, J.~J., \& {Stanford}, S.~A. 2003, \apj, 591, 749

\bibitem[{{Lucas} {et~al.}(2008){Lucas}, {Hoare}, {Longmore}, {Schr{\"o}der},
  {Davis}, {Adamson}, {Bandyopadhyay}, {de Grijs}, {Smith}, {Gosling},
  {Mitchison}, {G{\'a}sp{\'a}r}, {Coe}, {Tamura}, {Parker}, {Irwin}, {Hambly},
  {Bryant}, {Collins}, {Cross}, {Evans}, {Gonzalez-Solares}, {Hodgkin},
  {Lewis}, {Read}, {Riello}, {Sutorius}, {Lawrence}, {Drew}, {Dye}, \&
  {Thompson}}]{Lucas_2008}
{Lucas}, P.~W., {Hoare}, M.~G., {Longmore}, A., {et~al.} 2008, \mnras, 391, 136

\bibitem[{{Macri} {et~al.}(2019){Macri}, {Kraan-Korteweg}, {Lambert}, {Alonso},
  {Berlind}, {Calkins}, {Erdo{\u{g}}du}, {Falco}, {Jarrett}, \&
  {Mink}}]{Macri_2019}
{Macri}, L.~M., {Kraan-Korteweg}, R.~C., {Lambert}, T., {et~al.} 2019, \apjs,
  245, 6

\bibitem[{{Murgia} {et~al.}(2011){Murgia}, {Parma}, {Mack}, {de Ruiter},
  {Fanti}, {Govoni}, {Tarchi}, {Giacintucci}, \&
  {Markevitch}}]{2011A&A...526A.148M}
{Murgia}, M., {Parma}, P., {Mack}, K.~H., {et~al.} 2011, \aap, 526, A148

\bibitem[{{Oke} \& {Gunn}(1983)}]{1983ApJ...266..713O}
{Oke}, J.~B. \& {Gunn}, J.~E. 1983, \apj, 266, 713

\bibitem[{{Palmese} {et~al.}(2020){Palmese}, {Annis}, {Burgad}, {Farahi},
  {Soares-Santos}, {Welch}, {da Silva Pereira}, {Lin}, {Bhargava}, {Hollowood},
  {Wilkinson}, {Giles}, {Jeltema}, {Romer}, {Evrard}, {Hilton}, {Vergara
  Cervantes}, {Bermeo}, {Mayers}, {DeRose}, {Gruen}, {Hartley}, {Lahav},
  {Leistedt}, {McClintock}, {Rozo}, {Rykoff}, {Varga}, {Wechsler}, {Zhang},
  {Avila}, {Brooks}, {Buckley-Geer}, {Burke}, {Carnero Rosell}, {Carrasco
  Kind}, {Carretero}, {Castander}, {Collins}, {da Costa}, {Desai}, {De
  Vicente}, {Diehl}, {Dietrich}, {Doel}, {Flaugher}, {Fosalba}, {Frieman},
  {Garc{\'\i}a-Bellido}, {Gerdes}, {Gruendl}, {Gschwend}, {Gutierrez},
  {Honscheid}, {James}, {Krause}, {Kuehn}, {Kuropatkin}, {Liddle}, {Lima},
  {Maia}, {Mann}, {Marshall}, {Menanteau}, {Miquel}, {Ogando}, {Plazas},
  {Roodman}, {Rooney}, {Sahlen}, {Sanchez}, {Scarpine}, {Schubnell}, {Serrano},
  {Sevilla-Noarbe}, {Sobreira}, {Stott}, {Suchyta}, {Swanson}, {Tarle},
  {Thomas}, {Tucker}, {Viana}, {Vikram}, {Walker}, \& {DES
  Collaboration}}]{Palmese_2020}
{Palmese}, A., {Annis}, J., {Burgad}, J., {et~al.} 2020, \mnras, 493, 4591

\bibitem[{{Polletta} {et~al.}(2007){Polletta}, {Tajer}, {Maraschi},
  {Trinchieri}, {Lonsdale}, {Chiappetti}, {Andreon}, {Pierre}, {Le F{\`e}vre},
  {Zamorani}, {Maccagni}, {Garcet}, {Surdej}, {Franceschini}, {Alloin},
  {Shupe}, {Surace}, {Fang}, {Rowan-Robinson}, {Smith}, \&
  {Tresse}}]{Polletta_2007}
{Polletta}, M., {Tajer}, M., {Maraschi}, L., {et~al.} 2007, \apj, 663, 81

\bibitem[{{Rees}(1990)}]{1990MNRAS.244..233R}
{Rees}, N. 1990, \mnras, 244, 233

\bibitem[{{Schlafly} \& {Finkbeiner}(2011)}]{Schlafly_2011}
{Schlafly}, E.~F. \& {Finkbeiner}, D.~P. 2011, \apj, 737, 103

\bibitem[{{Schr{\"o}der} {et~al.}(2019{\natexlab{a}}){Schr{\"o}der},
  {Fl{\"o}er}, {Winkel}, \& {Kerp}}]{Schroeder_2019b}
{Schr{\"o}der}, A.~C., {Fl{\"o}er}, L., {Winkel}, B., \& {Kerp}, J.
  2019{\natexlab{a}}, \mnras, 489, 2907

\bibitem[{{Schr{\"o}der} {et~al.}(2019{\natexlab{b}}){Schr{\"o}der}, {van
  Driel}, \& {Kraan-Korteweg}}]{Schroeder_2019a}
{Schr{\"o}der}, A.~C., {van Driel}, W., \& {Kraan-Korteweg}, R.~C.
  2019{\natexlab{b}}, \mnras, 482, 5167

\bibitem[{{Skrutskie} {et~al.}(2006){Skrutskie}, {Cutri}, {Stiening},
  {Weinberg}, {Schneider}, {Carpenter}, {Beichman}, {Capps}, {Chester},
  {Elias}, {Huchra}, {Liebert}, {Lonsdale}, {Monet}, {Price}, {Seitzer},
  {Jarrett}, {Kirkpatrick}, {Gizis}, {Howard}, {Evans}, {Fowler}, {Fullmer},
  {Hurt}, {Light}, {Kopan}, {Marsh}, {McCallon}, {Tam}, {Van Dyk}, \&
  {Wheelock}}]{Skrutskie_2006}
{Skrutskie}, M.~F., {Cutri}, R.~M., {Stiening}, R., {et~al.} 2006, \aj, 131,
  1163

\bibitem[{{Soto} {et~al.}(2016){Soto}, {Lilly}, {Bacon}, {Richard}, \&
  {Conseil}}]{Soto_2016}
{Soto}, K.~T., {Lilly}, S.~J., {Bacon}, R., {Richard}, J., \& {Conseil}, S.
  2016, \mnras, 458, 3210

\bibitem[{{Stern} {et~al.}(2005){Stern}, {Eisenhardt}, {Gorjian}, {Kochanek},
  {Caldwell}, {Eisenstein}, {Brodwin}, {Brown}, {Cool}, {Dey}, {Green},
  {Jannuzi}, {Murray}, {Pahre}, \& {Willner}}]{Stern_2005}
{Stern}, D., {Eisenhardt}, P., {Gorjian}, V., {et~al.} 2005, \apj, 631, 163

\bibitem[{{Stone}(1996)}]{Stone_1996}
{Stone}, R. P.~S. 1996, \apjs, 107, 423

\bibitem[{{Strauss} {et~al.}(2002){Strauss}, {Weinberg}, {Lupton}, {Narayanan},
  {Annis}, {Bernardi}, {Blanton}, {Burles}, {Connolly}, {Dalcanton}, {Doi},
  {Eisenstein}, {Frieman}, {Fukugita}, {Gunn}, {Ivezi{\'c}}, {Kent}, {Kim},
  {Knapp}, {Kron}, {Munn}, {Newberg}, {Nichol}, {Okamura}, {Quinn}, {Richmond},
  {Schlegel}, {Shimasaku}, {SubbaRao}, {Szalay}, {Vanden Berk}, {Vogeley},
  {Yanny}, {Yasuda}, {York}, \& {Zehavi}}]{Strauss_2002}
{Strauss}, M.~A., {Weinberg}, D.~H., {Lupton}, R.~H., {et~al.} 2002, \aj, 124,
  1810

\bibitem[{{Valenti} {et~al.}(1998){Valenti}, {Piskunov}, \&
  {Johns-Krull}}]{Valenti_1998}
{Valenti}, J.~A., {Piskunov}, N., \& {Johns-Krull}, C.~M. 1998, \apj, 498, 851

\bibitem[{{van Weeren} {et~al.}(2011){van Weeren}, {R{\"o}ttgering}, \&
  {Br{\"u}ggen}}]{vanWeeren_2011}
{van Weeren}, R.~J., {R{\"o}ttgering}, H.~J.~A., \& {Br{\"u}ggen}, M. 2011,
  \aap, 527, A114

\bibitem[{{van Weeren} {et~al.}(2009){van Weeren}, {R{\"o}ttgering},
  {Br{\"u}ggen}, \& {Cohen}}]{vanWeeren_2009}
{van Weeren}, R.~J., {R{\"o}ttgering}, H.~J.~A., {Br{\"u}ggen}, M., \& {Cohen},
  A. 2009, \aap, 508, 75

\bibitem[{{Vazza} {et~al.}(2021){Vazza}, {Wittor}, {Brunetti}, \&
  {Br{\"u}ggen}}]{2021arXiv210204193V}
{Vazza}, F., {Wittor}, D., {Brunetti}, G., \& {Br{\"u}ggen}, M. 2021, arXiv
  e-prints, arXiv:2102.04193

\bibitem[{{Warren} {et~al.}(2007){Warren}, {Cross}, {Dye}, {Hambly}, {Almaini},
  {Edge}, {Hewett}, {Hodgkin}, {Irwin}, {Jameson}, {Lawrence}, {Lucas},
  {Mortlock}, {Adamson}, {Bryant}, {Collins}, {Davis}, {Emerson}, {Evans},
  {Gonzales-Solares}, {Hirst}, {Kerr}, {Lewis}, {Mann}, {Rawlings}, {Read},
  {Riello}, {Sutorius}, \& {Varricatt}}]{Warren_2007}
{Warren}, S.~J., {Cross}, N.~J.~G., {Dye}, S., {et~al.} 2007, arXiv e-prints,
  astro

\bibitem[{{Wechsler} \& {Tinker}(2018)}]{Wechsler_2018}
{Wechsler}, R.~H. \& {Tinker}, J.~L. 2018, \araa, 56, 435

\bibitem[{{Wen} \& {Han}(2013)}]{Wen_2013}
{Wen}, Z.~L. \& {Han}, J.~L. 2013, \mnras, 436, 275

\bibitem[{{Werner} {et~al.}(2004){Werner}, {Roellig}, {Low}, {Rieke}, {Rieke},
  {Hoffmann}, {Young}, {Houck}, {Brandl}, {Fazio}, {Hora}, {Gehrz}, {Helou},
  {Soifer}, {Stauffer}, {Keene}, {Eisenhardt}, {Gallagher}, {Gautier}, {Irace},
  {Lawrence}, {Simmons}, {Van Cleve}, {Jura}, {Wright}, \&
  {Cruikshank}}]{Werner_2004}
{Werner}, M.~W., {Roellig}, T.~L., {Low}, F.~J., {et~al.} 2004, \apjs, 154, 1

\bibitem[{{Ziparo} {et~al.}(2016){Ziparo}, {Smith}, {Mulroy}, {Lieu}, {Willis},
  {Hudelot}, {McGee}, {Fotopoulou}, {Lidman}, {Lavoie}, {Pierre}, {Adami},
  {Chiappetti}, {Clerc}, {Giles}, {Maughan}, {Pacaud}, \&
  {Sadibekova}}]{Ziparo_2016}
{Ziparo}, F., {Smith}, G.~P., {Mulroy}, S.~L., {et~al.} 2016, \aap, 592, A9

\end{thebibliography}
%\bibliography{test} % your references Yourfile.bib

\end{document}